\documentclass[acmsmall,authorversion]{acmart}

\usepackage{amsmath}
\usepackage{bm}
\usepackage{xspace}
\usepackage{lipsum}
\usepackage{float}
\usepackage{graphicx}
\usepackage{multirow}
\usepackage{subcaption}
\usepackage{siunitx}
\usepackage{tabularx}
\newcommand{\tablefontsize}{\small}
\newcommand{\tablenotesfontsize}{\small}
\newcommand{\tableheadline}{\textbf}
\usepackage{colortbl}
\newcommand{\altrowcolor}{gray!10}
\usepackage{enumitem}

\newlist{tableitemize}{itemize}{1}
\setlist[tableitemize,1]{label={--},nosep, leftmargin=*, topsep=0pt, partopsep=0pt, before=\vspace{-0.5\baselineskip}, after=\vspace{0.1\baselineskip}}

\newcommand{\tnotex}{\textsuperscript}
\newcommand{\tablenotesx}[2]{\multicolumn{#1}{p{0.95\columnwidth}}{\tablenotesfontsize{#2}}}

\newcommand{\sig}[1]{\tnotex{\pnote}\makebox[-3pt][r]{\textbf{#1}}}
\newcommand{\ssig}[1]{\tnotex{\ppnote}\makebox[-5pt][r]{\textbf{#1}}}

\newcommand{\briefSectionBF}[1]{\medskip\noindent\textbf{{#1}.}}

\newcommand{\rating}[1]{\noindent$\langle$\textit{{#1}}$\rangle$}
\newenvironment{task}{}{}

\newcommand{\SafeSlinger}{SafeSlinger\xspace}
\newcommand{\OurApproach}{PairSonic\xspace}

\newcommand{\pnote}{*}
\newcommand{\ppnote}{**}

\usepackage[english]{babel} % enables calling "Subsubsections" "Section"
\usepackage{babel}
\addto\extrasenglish{%

}
\microtypecontext{spacing=nonfrench}

\usepackage[acronym,nohypertypes={acronym},shortcuts,nonumberlist,nolist]{glossaries} % all acronyms
\newacronym{AAS}{AAS}{Account Adoption Scale}
\newacronym{ATI}{ATI}{Affinity for Technology Interaction}
\newacronym{BLE}{BLE}{Bluetooth Low Energy}
\newacronym{COTS}{COTS}{commercial off-the-shelf}
\newacronym{E2EE}{E2EE}{end-to-end encrypted}
\newacronym{CTAP}{CTAP}{Client to Authenticator Protocol}
\newacronym[description={Deutsches Forschungsnetz ("German research network")}]{DFN}{DFN}{Deutsches Forschungsnetz}
\newacronym{DNS}{DNS}{Domain Name System}
\newacronym{FIDO}{FIDO}{Fast IDentity Online Alliance}
\newacronym{FSK}{FSK}{Frequency-Shift Keying}
\newacronym{GDPR}{GDPR}{General Data Protection Regulation}
\newacronym{HTTPS}{HTTPS}{Hypertext Transfer Protocol Secure}
\newacronym{IP}{IP}{Internet Protocol}
\newacronym{ISO}{ISO}{International Organization for Standardization}
\newacronym{NIST}{NIST}{National Institute of Standards and Technology}
\newacronym{OTSP}{OTSP}{Online Token Status Protocol}
\newacronym{PAS}{PAS}{Participant Adoption Scale}
\newacronym{SMS}{SMS}{Short Message Service}
\newacronym{SUS}{SUS}{System Usability Scale}
\newacronym{TLS}{TLS}{Transport Layer Security}
\newacronym{TUDA}{TUDA}{Technische Universität Darmstadt}
\newacronym{U2F}{U2F}{Universal 2nd Factor}
\newacronym{UAF}{UAF}{Universal Authentication Framework}
\newacronym{USB}{USB}{Universal Serial Bus}
\newacronym{W3C}{W3C}{World Wide Web Consortium}
\newacronym{WebAuthn}{WebAuthn}{Web Authentication}

\newacronym[description={One-Factor Authentication}]{1FA}{1FA}{one-factor authentication}
\newacronym[description={Two-Factor Authentication}]{2FA}{2FA}{two-factor authentication}
\newacronym[description={Application Programming Interface}]{API}{API}{application programming interface}
\newacronym[description={Application}]{app}{app}{application}
\newacronym[description={Certification Authority}]{CA}{CA}{certification authority}
\newacronym[description={Effect Size}]{ES}{ES}{effect size}
\newacronym[description={False Discovery Rate}]{FDR}{FDR}{false discovery rate}
\newacronym[description={False Rejection Rate}]{FRR}{FRR}{false rejection rate}
\newacronym[description={Graphical User Interface}]{GUI}{GUI}{graphical user interface}
\newacronym[description={Institutional Review Board}]{IRB}{IRB}{institutional review board}
\newacronym[description={Internet Service Provider}]{ISP}{ISP}{internet service provider}
\newacronym[description={Multi-Factor Authentication}]{MFA}{MFA}{multi-factor authentication}
\newacronym[description={Machine-in-the-Middle}]{MitM}{MitM}{machine-in-the-middle}
\newacronym[description={Near-Field Communication}]{NFC}{NFC}{near-field communication}
\newacronym[description={Operation System}]{OS}{OS}{operating system}
\newacronym[description={One-Time Password}]{OTP}{OTP}{one-time password}
\newacronym[description={Out-Of-Band}]{OOB}{OOB}{out-of-band}
\newacronym[description={Personal Computer}]{PC}{PC}{personal computer}
\newacronym[description={Personal Identification Number}]{PIN}{PIN}{personal identification number}
\newacronym[description={Public-Key Infrastructure}]{PKI}{PKI}{public-key infrastructure}
\newacronym[description={Quick-Response}]{QR}{QR}{quick-response}
\newacronym[description={Secure Device Pairing}]{SDP}{SDP}{secure device pairing}
\newacronym[description={Service Set Identifier}]{SSID}{SSID}{service set identifier}
\newacronym[description={Single Sign-On}]{SSO}{SSO}{single sign-on}
\newacronym[description={Transaction Authentication Number}]{TAN}{TAN}{transaction authentication number}
\newacronym[description={Trust on First Use}]{TOFU}{TOFU}{trust on first use}
\newacronym[description={Trusted Platform Module}]{TPM}{TPM}{trusted platform module}
\newacronym[description={Uniform Resource Locator}]{URL}{URL}{uniform resource locator}
\newacronym[description={Wireless Local Area Network}]{WLAN}{WLAN}{wireless local area network}

\newcommand\VRule[1][\arrayrulewidth]{\vrule width #1}
\newcommand{\studyquote}[2]{{\def\arraystretch{2}\setlength\tabcolsep{7pt}\vspace{1ex} \noindent\begin{tabular}{!{\color{gray!75}\VRule[2pt]}p{\dimexpr\linewidth-2\tabcolsep-0.3pt}}\cellcolor{gray!10}\textit{``#1''} \mbox{{(Group #2)}}\tabularnewline\end{tabular}\vspace{1ex}}}

\newcommand{\textquote}[1]{\textit{``#1''}}

\AtBeginDocument{%
  \providecommand\BibTeX{{%
    \normalfont B\kern-0.5em{\scshape i\kern-0.25em b}\kern-0.8em\TeX}}}

\microtypecontext{spacing=nonfrench}

\setcopyright{acmlicensed}
\acmJournal{PACMHCI}
\acmYear{2024} \acmVolume{8} \acmNumber{CSCW2}
\acmArticle{425} \acmMonth{11} \acmPrice{15.00}
\acmDOI{10.1145/3686964}

\received{July 2023}
\received[revised]{January 2024}
\received[accepted]{March 2024}

\acmSubmissionID{1160}

\begin{document}

\title[Sounds Good? Fast and Secure Contact Exchange in Groups]{Sounds Good? Fast and Secure Contact Exchange in Groups}

\author{Florentin Putz}
\orcid{0000-0003-3122-7315}
\affiliation{%
  \institution{Technical University of Darmstadt}
  \city{Darmstadt}
  \country{Germany}
  }
\email{fputz@seemoo.de}

\author{Steffen Haesler}
\orcid{0000-0002-6808-0487}
\affiliation{%
    \institution{Technical University of Darmstadt}
    \city{Darmstadt}
    \country{Germany}
    }
\email{haesler@peasec.tu-darmstadt.de}

\author{Matthias Hollick}
\orcid{0000-0002-9163-5989}
\affiliation{%
  \institution{Technical University of Darmstadt}
  \city{Darmstadt}
  \country{Germany}
}
\email{mhollick@seemoo.de}

\renewcommand{\shortauthors}{Florentin Putz, Steffen Haesler, and Matthias Hollick}

\begin{abstract}
Trustworthy digital communication requires the secure  exchange of contact information, but current approaches lack usability and scalability for larger groups of users.
We evaluate the usability of two secure contact exchange systems: the current state of the art, SafeSlinger, and our newly designed protocol, \textit{\OurApproach},
which extends trust from physical encounters to spontaneous online communication.
Our lab study ($N=45$) demonstrates \OurApproach's superior usability, automating the tedious verification tasks from previous approaches via an acoustic out-of-band channel.
Although participants significantly preferred our system, minimizing user effort surprisingly decreased the perceived security for some users, who associated security with complexity.
We discuss user perceptions of the different protocol components and identify remaining usability barriers for CSCW application scenarios.
\end{abstract}



\begin{teaserfigure}
  \centering
  \includegraphics[width=0.7\textwidth]{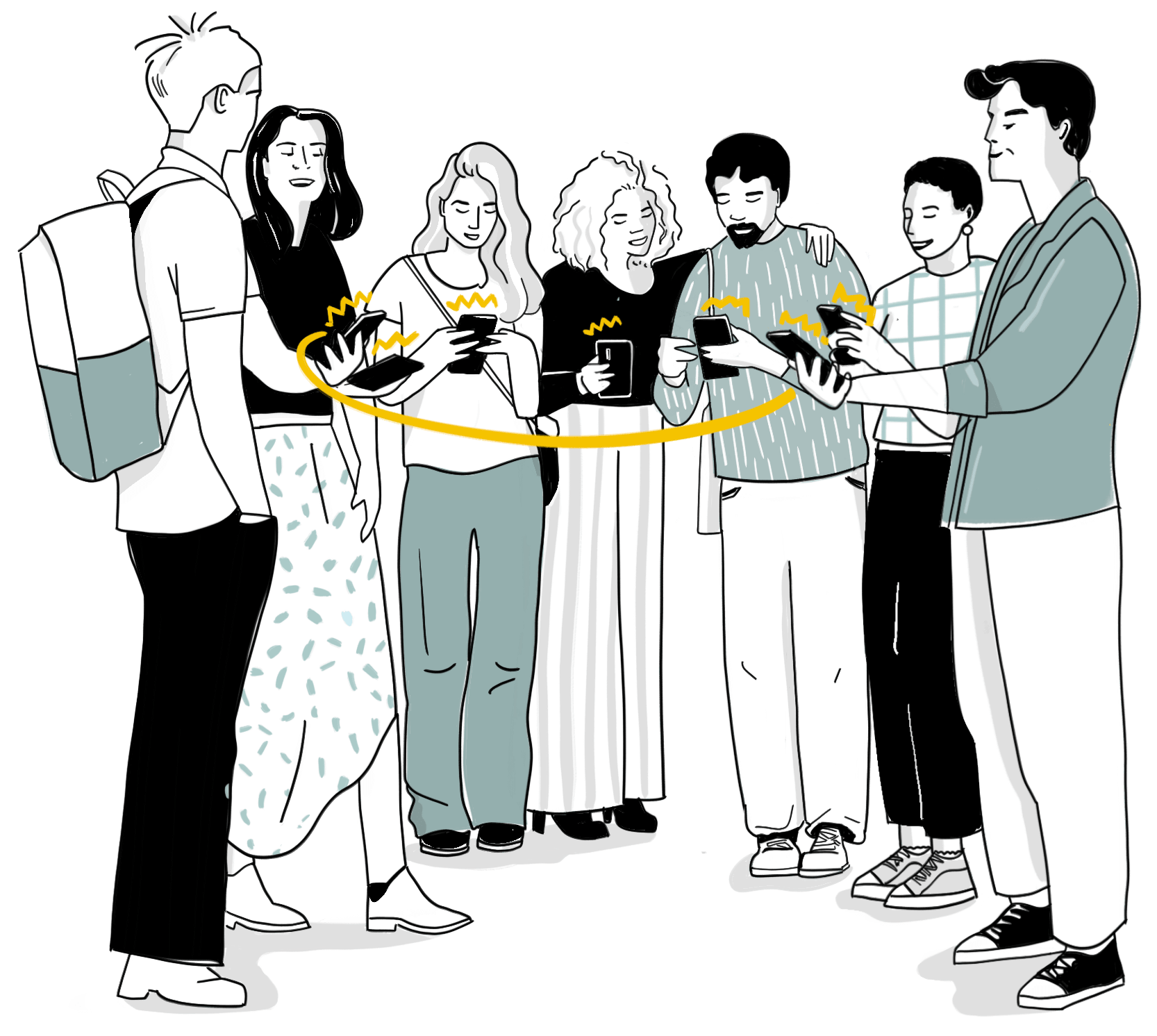}
  \caption{Group of users exchanging their contact information using our acoustic group pairing protocol.}
  \Description{Illustration of a group of seven people standing in a circle, each holding a smartphone. They are exchanging contact information using the PairSonic protocol, depicted by lines connecting the phones, symbolizing data transmission. The environment suggests an informal, collaborative setting. This figure emphasizes the group dynamic and interaction enabled by PairSonic.}
  \label{fig:teaser}
\end{teaserfigure}

\maketitle
\section{Introduction}
Imagine a group of seven researchers who just met for the first time at a conference.
During the event, they notice that they share a lot of common interests -- they might even have some ideas for future collaborations -- so they decide to stay in contact.
They all have a smartphone, but to securely communicate online, they need a way to \emph{securely exchange their contact information}.
In this paper, we study \textit{group pairing} methods for the fast and secure exchange of contact information that fulfill the following two goals:
first, they allow the participants to contact each other online, to establish new conversations or to start new digital collaborations.
Second, they authenticate the participants by leveraging their physical encounter to verify their contact information.\footnote{The exchanged contact information usually contains additional metadata and public-key material, e.g., for secure \gls{E2EE} instant messaging.}

There are countless similar use cases for online collaboration besides the aforementioned group of researchers, ranging from professional activities (team communication, remote work, project management), to educational settings (interactive learning platforms, knowledge management), and leisure (event planning and coordination), affecting billions of users in today's digital world \cite{studer2008Survey,kindberg2002System,lundgren2015Designing,handel2002What,namara2021differential}.
For all these scenarios, secure exchange of contact information is required to associate known physical individuals with their online profiles and protect their privacy and security when interacting online.
Without such a secure contact exchange, an adversary can eavesdrop on private communication or even impersonate one of the communication participants in order to modify or create fake messages that appear legitimate.

Effective communication and collaboration thrives in a trusted environment, where participants have clear knowledge of the identities of others involved and no concerns about unauthorized access to their conversations.
Such an environment encourages open discussions, even on sensitive topics \cite{shusas2023accounting}.
Previous research has shown that users desire authenticated conversations -- wanting to know exactly \emph{who} they are communicating with -- when talking to friends and family members \cite{fassl2023Why}, but also when communicating with colleagues or business partners \cite{fassl2021Exploring}, and especially so for vulnerable groups such as journalists and activists \cite{herzberg2021Secure}.
Secure contact exchange is therefore an important prerequisite for fruitful and trustworthy online communication and collaboration \cite{oesch2022User}.

\subsection{Current Group Pairing Is Insufficient}\label{sec:current-pairing-insufficient}
Throughout the last decade, users have increasingly adopted \acrfull{E2EE} tools such as Signal \cite{signalMessenger}, Whatsapp \cite{whatsappMessenger}, Telegram \cite{telegramMessenger}, and Threema \cite{threemaMessenger}, facilitating group collaboration with features like real-time messaging, file sharing, and video calls.
These \gls{E2EE} tools also offer the verification of contact information through an \emph{authentication ceremony}, which requires users to perform a manual verification action, either by scanning \gls{QR} codes or by manually comparing public key fingerprints on their devices.
This authentication ceremony allows users to transfer an existing trust relationship from a physical encounter to the corresponding digital identity of their communication partners, enabling authenticated conversations.
There are two problems with the design of authentication ceremonies in state-of-the-art tools for secure communication and collaboration:

\begin{enumerate}
    \item \textbf{Low usability.} Current authentication ceremonies are \emph{hard to use}, even for professionals.
    \item \textbf{Bad scalability.} Current authentication ceremonies \emph{do not support larger groups} of users.
\end{enumerate}

First, previous usability research has shown that users struggle to understand and perform authentication ceremonies \cite{oesch2022User,vaziripour2017that,vaziripour2018Action,wu2019Something}. 
Most users expect that \gls{E2EE} tools like Signal are secure by default without further verification actions, when in reality these tools cannot protect against active \gls{MitM} attacks and rogue service operators unless the users diligently perform the authentication ceremony \cite{herzberg2021Secure,vasile2020ghost}.
These authentication ceremonies, however, require considerable user interaction \cite{herzberg2016Can}, take a long time \cite{vaziripour2017that}, and are difficult to use even for security professionals \cite{schroder2016When,fassl2023Why}.
Recently, the risks of active \gls{MitM} attacks in \gls{E2EE} tools have gained more attention due to reports of law enforcement agencies compromising app providers \cite{levy2018principles,goodin2018police}, demonstrating a practical and concerning risk for privacy infringements, especially if such vulnerabilities were to be exploited by malicious actors in the future.

Second, 
state-of-the-art tools like Signal and WhatsApp do not support efficient and secure contact exchange for \emph{groups of multiple users}, but only between two users.
This is a substantial shortcoming, as collaboration often involves more than two participants.
While each pair of group members could separately perform the bilateral authentication ceremony one after another, this is infeasible as the number of manual verification actions scales quadratically ($\frac{n(n-1)}{2}$) with the group size ($n$).
To illustrate this:
if our exemplary group of seven researchers were to use Signal, they would need to perform 21 manual actions amongst each other for authentication, but each user can only perform one of these actions at a time. Such an authentication takes too much time and would discourage the spontaneous decision to collaborate.

The main challenge is how to authenticate the exchanged contact information and key material without having to rely on any trusted third parties or \glspl{PKI} that facilitate the exchange.
Farb et al.~\cite{farb2013SafeSlinger} achieved a breakthrough towards solving this problem when they proposed the more usable \textit{SafeSlinger} protocol for the secure exchange of contact information in the absence of trusted third parties.
While SafeSlinger offers significantly higher usability than alternatives~\cite{farb2013SafeSlinger}, it still requires the participants to perform $\frac{n(n-1)}{2}$ manual text comparisons to achieve its security guarantees.
Since the release of SafeSlinger, several studies on authentication methods suggested that users generally prefer simpler schemes requiring less involvement \cite{xu2021Key,mehrabi2019Evaluating} and that involving users for manual comparisons is prone to errors \cite{kainda2009Usability,tan2017Can,vaziripour2017that}.
Furthermore, recent research has explored the intuitive actions users might take to associate multiple devices \cite{chong2013How,jokela2015Connecting}, suggesting that the user experience in SafeSlinger may not align with user expectations.
The required amount of user interaction is potentially discouraging, especially considering larger groups.
Instead of facilitating trustworthy communication and collaboration, state-of-the-art methods rather inhibit it (see \autoref{sec:related-work} for a more extensive discussion of related work).

\subsection{Approach}
In this work, we approach the challenge of securely exchanging contact information from a different angle, under the research hypothesis that \emph{users would prefer a contact exchange method with minimal user interaction} \cite{kuo2008Mind,xu2021Key,mehrabi2019Evaluating}.
Consequently, we design a new group pairing protocol named \textit{\OurApproach} that requires less effort and can efficiently scale to multiple participants.
The illustrative group of researchers can exchange their contact information securely and quickly by starting \OurApproach and simply holding their devices close together for a few seconds, aligning with intuitive association behavior \cite{chong2013How,jokela2014FlexiGroups,jokela2015Connecting}.
\OurApproach automatically exchanges the
participants' contact information using an ad-hoc WiFi channel for communication and a location-limited acoustic \gls{OOB} channel for verification.
The effectiveness of the acoustic channel has already been demonstrated in previous research \cite{mehrabi2019Evaluating,gerasimovThingsThatTalk2000,madhavapeddy2003ContextAware,lopesAcousticModemsUbiquitous2003} and in industry adoption, such as with the Sonos smart speaker \cite{sonos2022near}, for the similar use case of pairwise key verification.
Even though the acoustic channel seems promising, its applicability and usability between more than two devices remains unexplored.

\subsection{Contributions}
Our main contribution to CSCW is the usability evaluation of our novel acoustic approach \OurApproach for the secure exchange of contact information and cryptographic public keys for groups of multiple users as a basic requirement for teams to work with sensitive content.
We conduct a lab study with $N=45$ participants -- consisting of practical authentication tasks on smartphones, subsequent interviews, and a questionnaire -- to compare the state-of-the-art system SafeSlinger with our novel protocol \OurApproach, which we design and implement for smartphones.
\OurApproach requires only minimal user interaction by leveraging an acoustic \gls{OOB} channel between the users' smartphones, automating the cumbersome verification tasks from previous commercial and academic approaches \cite{vaziripour2017that,herzberg2021Secure,unger2015SoK,dechand2016Empirical,uzun2007Usability}.
This paper is structured as follows:

\begin{itemize}
    \item We define the problem and establish our requirements for secure contact exchange (\autoref{sec:problem}), discussing why the state of the art fails to solve this problem (\autoref{sec:related-work}).
    \item We design \OurApproach, an improved protocol  for fast and secure contact exchange, and implement it for smartphones (\autoref{sec:design}).
    \item To answer our research questions (\autoref{sec:research-questions}),
    aiming to understand how automating the cumbersome verification can improve on the current state of the art (\SafeSlinger),
    we conduct a lab study with $N=45$ participants to compare both systems (\autoref{sec:methods}).
    \item Based on the quantitative insights from our questionnaire (\autoref{sec:quantitative}) and qualitative interviews (\autoref{sec:qualitative}), we discuss our observation that less user interaction seems to improve usability in our case, but can have unexpected negative effects on perceived security. We evaluate the suitability of the acoustic \gls{OOB} channel for pairing and give recommendations how to improve the state of the art of secure contact exchange (\autoref{sec:discussion}).
    \item We provide a replication package with our evaluation scripts and the pseudonymized dataset from our study, which contains 69 variables for each of the 45 participants \cite{zenodo2024dataset}.
\end{itemize}

\section{Problem Setting}\label{sec:problem}
This section defines the problem we are addressing (\autoref{sec:problem-definition}) and our assumptions (\autoref{sec:assumptions}).
\subsection{Problem Definition}\label{sec:problem-definition}
Our goal is to let a group of two or more users that physically meet (\autoref{fig:teaser}) spontaneously exchange their contact information \cite{farb2013SafeSlinger}.
The expected outcome of this \emph{secure contact exchange}\footnote{
The process of exchanging and verifying the identities with a group of participants is sometimes also called \emph{group key verification}, or \emph{group authentication ceremony}, or \emph{peer-to-peer authentication}, or \emph{secure group contact sharing}, or \emph{key establishment for groups}, or \emph{spontaneous device association}.
In this work, we call this process \emph{secure contact exchange} or \emph{group pairing}, as we always target groups of two or more users unless otherwise noted.
}
is that each participant obtains the authentic and verified contact information of all other participants, including their cryptographic public keys, without having to rely on external key management infrastructure, prior associations, or shared secrets.
We target a group of users who physically meet for the pairing process (\autoref{sec:discussion-meetings}) and want to be able to set up secure communication channels between all subsets of users within the group.
After this pairing process, the users can subsequently communicate in a confidential and authenticated manner over untrusted channels such as the Internet.
Our contact exchange protocol has the following requirements:

\medskip
\noindent\textsc{Usability Requirements}
\begin{enumerate}
    \item \textbf{Minimal user interaction}, due to our research hypothesis that users would prefer such a contact exchange method \cite{kuo2008Mind,xu2021Key,mehrabi2019Evaluating}.
    The protocol should be easy to use for lay users, facilitating spontaneous ad-hoc collaboration plans \cite{chong2012Usability,chong2014Survey}.
    \item \textbf{Scalable to larger groups of users.} The pairing protocol should be fast and not noticeably depend on the number of users.
    \item \textbf{Error-proof,} as users can make mistakes. In particular, our protocol should be robust to \emph{rushing users} who prioritize speed and efficiency over adhering to instructions or prompts, potentially compromising the security of the protocol.
\end{enumerate}

\medskip
\noindent\textsc{Deployability Requirements}
\begin{enumerate}\setcounter{enumi}{3}
    \item \textbf{Compatibility.} Our protocol should seamlessly run on existing \gls{COTS} devices, enabling users to leverage its benefits immediately by simply installing our software. We focus on user-controlled wireless devices such as smartphones, tablets, laptops, and smart watches, without requiring further hardware or firmware modifications.
    \item \textbf{Ad-hoc.} Our protocol should not require any existing security context in the form of pre-shared keys or a jointly trusted third party.
    \item \textbf{Decentral.}
    Our protocol should operate independently of any central infrastructure and remain fully functional without an active Internet connection.
    This requirement is crucial to support people's spontaneous decision-making, facilitating seamless collaboration and communication regardless of their location.
    This means that no metadata can leak to third parties, as part of our privacy-by-design approach.
    If a security vulnerability arises in our protocol or implementation, the offline nature of our exchange makes exploitation by attackers significantly more challenging.
\end{enumerate}

\medskip
\noindent\textsc{Security Requirements}
\begin{enumerate}\setcounter{enumi}{6}
    \item \textbf{Confidentiality.} Only the intended participants should receive the contact information.
    \item \textbf{Contact authentication.} Each received contact information should match the corresponding honest participant. We require mutual authentication, i.e., each participant authenticates all the other participants.
    \item \textbf{Collective pairwise security.} This requirement was postulated by Farb et al.~\cite{farb2013SafeSlinger}, stipulating that each pair of participants in our groupwise exchange should get the same security properties as if they had performed a direct pairwise exchange. This implies that adversarial participants cannot degrade the bindings between other honest participants. 
\end{enumerate}

\subsection{Assumptions}\label{sec:assumptions}
We assume that all users are physically present at the same location and can utilize the concept of \textquote{group demonstrative identification} \cite{li2010Group}, which involves legitimate participants excluding unintended participants or imposters based on personal identification, such as appearance or voice.
The denotation of intended and legitimate participants is an inherently social classification -- the decision of whom \textit{users} expect to communicate with cannot be achieved by technical measures alone and needs human assistance.
Users are responsible for verifying the received contact information from their exchange to ensure it matches their intended participants \cite{farb2013SafeSlinger}. 
They should also reject non-participants and detect any instances of impersonation within the group, such as duplicate entries corresponding to honest participants.
We also assume that the participants accurately determine their group size.

\subsection{Adversary Model}\label{sec:adversary-model}
Although a detailed security evaluation is beyond the scope of this work, we consider potential threats from an adversary with diverse capabilities. The attacker's goal is to breach the security requirements stated in \autoref{sec:problem-definition}. For example, the adversary might attempt to impersonate one of the legitimate participants by making another participant accept an adversarial cryptographic public key. Such an attack would allow the attacker to launch \gls{MitM} attacks in the future, manipulating or intercepting confidential communication.
However, ensuring the uninterrupted availability of the contact exchange process is not one of our security goals. Therefore, if an adversary disrupts the contact exchange (e.g., by jamming the radio channel), users would need to retry the exchange at another time or place.

Our adversary model is based on the model used in SafeSlinger \cite{farb2013SafeSlinger}, focusing on both nearby and remote adversaries.
These adversaries possess full Dolev-Yao \cite{dolev1983security} control over WiFi network messages, meaning they can eavesdrop, intercept, modify, replay, or inject radio communication between devices. Likewise, we also assume that the adversary has not compromised the hardware and software of the participants' smartphones, as such attacks are beyond this paper's scope.

In addition to SafeSlinger's adversary model, our research also considers the acoustic out-of-band channel. Based on previous research by Stajano and Anderson \cite{stajano2000Resurrecting} and Balfanz et al.~\cite{balfanz2002Talking}, we regard this channel as location-limited, characterized by its ability to support \textit{demonstrative identification} and ensure \textit{authenticity} due to the inherent physical constraints of acoustic sound pressure waves.
Consequently, while adversaries can eavesdrop on this channel, they cannot transmit undetected. We explore this assumption and potential attacks on the acoustic channel in more detail in \autoref{sec:discussion-security}.

\section{Related Work}\label{sec:related-work}
Our work focuses on the problem of securely exchanging contact information, including cryptographic key material, in groups of users.
This problem relates to several branches of research, which we first summarize and then elaborate in further detail:

\begin{itemize}
    \item \textbf{Device association}\footnote{Device association is sometimes also called binding, coupling, or bonding \cite{chong2014Survey}.} is the general act of establishing a communication channel between two or more devices, independent of the duration and the association's security \cite{chong2014Survey}. Our problem is a special case of device association with a specific focus on authenticity and security to facilitate trustworthy communication.
    \item \textbf{Secure device pairing} is closely related to our work, as it involves device association aimed at establishing an authentic communication channel \cite{fomichevSurveySystematizationSecure2018}. Whereas previous work mostly focused on connecting two devices (\textquote{pair\emph{ing}}), our work additionally considers the more complex case of connecting multiple users, although it also functions with just two users.
    A key distinction is that we aim not only to establish a secure communication channel for the entire group, but potentially for every subset of group members as well: CSCW research has shown that larger groups sometimes like to split dynamically into smaller task forces for more effective collaboration \cite{harandi2019supporting,umbelino2019prototeams}. The group exists temporarily to facilitate the exchange of contact information among its members, but subsequent communication can happen independently of the group's context. Naturally, the group members can also opt to collaborate as a whole group.
    \item \textbf{Acoustic communication} can be used as a physical \gls{OOB} channel to transmit data (such as contact information) between nearby devices, which is compatible with all smart devices having a microphone and a speaker.
    Previous work showed promising results for connecting two devices \cite{gerasimovThingsThatTalk2000,lopesAcousticModemsUbiquitous2003,mehrabi2019Evaluating}. We are the first, to the best of our knowledge, to study the suitability and usability of acoustic communication for connecting multiple users.
    \item \textbf{Authentication ceremonies} refer to the process by which users can verify their cryptographic keys in the context of \gls{E2EE} tools \cite{herzberg2016Can}.
    Existing approaches only have limited usability and do not support more than two users. In contrast, our approach aims to be a more usable authentication ceremony supporting multiple participants.
\end{itemize}

\subsection{Device Association}
Device association methods establish a basic communication channel between two or more devices \cite{chong2014Survey,jokela2013comparative,jokela2014FlexiGroups}.
However, since these methods do not consider the security of the association, they do not directly aid secure contact exchange.
Nonetheless, usability evaluations exist on preferred user actions or gestures for associating their device with others \cite{kray2010Userdefined,gronbaek2020Proxemics,chong2012Usability}.
These evaluations can help us optimize the user interaction for the secure exchange of contact information, which also requires temporary device association with other participants.
One such line of work is especially interesting: guessability studies with plastic surrogates to determine intuitive user actions for associating a group of devices \cite{chong2011How,chong2013How,jokela2015Connecting,jokela2016Natural}.
Instead of multiple pairwise interactions, users preferred a singular group-wise interaction, favorably by pointing their smartphones towards each other or bringing them into proximity.
These results inform the design of our approach, where we use a single group-wise interaction, requiring all smartphones to be in close proximity during the exchange.

\subsection{Secure Device Pairing}\label{sec:relatedwork-sdp}
There is a large body of work dealing with the secure exchange of cryptographic public keys, called secure device pairing, which is very similar to our more general problem of the secure exchange of contact information.
This line of work mostly focused on establishing keys between two devices \cite{fomichevSurveySystematizationSecure2018}, and previous usability studies focused mostly on this use case \cite{uzun2007Usability,kumar2009comparative,kobsa2009Serial,kainda2009Usability,uzun2011Pairing,ion2010Influence,chong2012Usability}. Pairing between two devices does not securely generalize to more than two parties due to new types of attacks that are possible when there are multiple participants \cite{kuo2008Mind,farb2013SafeSlinger}.

However, there is also some prior work on establishing keys between multiple devices, which roughly falls into one of four categories: (1) human-in-the-loop, (2) shared homogeneous context, (3) physical layer security, (4) \gls{OOB} channels.

\subsubsection{Humans-in-the-loop}
There is previous research focusing on key establishment where each participant has to perform some manual action, such as entering a shared secret \cite{burmester1997Efficient,abdalla2006PasswordBased,asokan2000Key,valkonen2006Ad}, comparing text \cite{valkonen2006Ad,laur2008SASBased,nguyen2008Authenticating,mezzour2010HoPo,studer2008Survey} or visual patterns \cite{keoh2009Securing,chenGAnGSGatherAuthenticate2008,linSPATESmallgroupPKIless2009,linSPATESmallGroupPKILess2010}, or moving all devices in a correlated manner \cite{mayrhofer2009Shake,kirovski2007Martini}.
Besides the obvious drawback that these schemes require substantial user interaction and do not scale well, they often cannot protect against malicious insiders \cite{kuo2008Mind}.

\subsubsection{Shared Homogeneous Context (Zero-interaction Pairing)}
Zero-interaction schemes try to establish a shared key based on the entropy of observable common events in the surroundings \cite{xu2021Key,fomichev2019perils}.
While such approaches at first glance seem to fit well to our requirement of minimal user interaction, they are not suitable for user-controlled device association and take more time than is tolerable for an ad-hoc spontaneous exchange \cite{li2020T2Pair,farrukh2023One}.
These schemes were designed for a different use case, namely for pairing multiple devices owned by a single user, where the key establishment process is allowed to take its time in the background, without user involvement.
Instead, group pairing is a social use case, which not only involves devices but also multiple people, and thus a group pairing protocol should not neglect the social interaction in the group.

\subsubsection{Physical Layer Security}
Another approach makes use of the physical laws of signal propagation on the wireless radio channel to securely establish keys \cite{houChorusScalableInband2013,ghose2017HELP,ghose2018Secure,ghose2022InBand,han2021DroneKeya}. A similar line of work uses a Faraday cage to prevent outside attackers from interacting with the key establishment \cite{kuo2007Messageinabottle,law2011KALwEN}.
These approaches are often hard to deploy, as they either require specific firmware or hardware, making them incompatible with \gls{COTS} devices, or they require additional devices to assist with the pairing process.

\subsubsection{Out-of-Band Channels}
One line of work studies key establishment using \gls{OOB} channels, which are low-bandwidth auxiliary channels that are usually location-limited to protect the integrity of its messages \cite{stajano2000Resurrecting}.
Previous research considered infrared channels \cite{balfanz2002Talking}, light channels \cite{saxena2009Blink,perkovic2012Secure,kovacevic2016Flashing}, and \gls{NFC} \cite{chongSpatialColocationDevice2011}.
The audio channel has also been explored as a promising \gls{OOB} channel for pairing two devices \cite{soriente2008HAPADEP,goodrich2006Loud,goodrich2009Using,putz2020AcousticIntegrityCodes}.
As we develop this concept further for pairing multiple devices, we will elaborate on the details of the audio channel in the following section.

\subsection{Acoustic Communication}\label{sec:related-acousticcommunication}
The acoustic channel is an attractive choice for connecting nearby devices, thanks to its ubiquitous compatibility with smartphones by repurposing their integrated audio hardware \cite{gerasimovThingsThatTalk2000}.
The physical layer can be entirely software-defined \cite{lopesAcousticModemsUbiquitous2003},
facilitating easy adaptation of modem properties to fit specific communication scenarios.
Audio communication is often employed as a location-limited \gls{OOB} channel, as it has much shorter range than radio communication and is restricted by physical barriers like walls \cite{soriente2008HAPADEP}.
Several proposals in the literature have explored using audio hardware for communication, including both audible and ultrasonic inaudible frequencies \cite{nandakumarDhwaniSecurePeertopeer2013,santagatiUWearSoftwareDefinedUltrasonic2015,wangMessagesSound2016,lee2015Chirp,getreuerUltrasonicCommunicationUsing2018,putz2020AcousticIntegrityCodes}.

We are only aware of one previous study by Mehrabi et al.~\cite{mehrabi2019Evaluating} evaluating the usability of acoustic data transmission.
In a lab study involving 12 participants in a pairwise data exchange scenario, they compared audible acoustic communication with \gls{BLE} and \gls{QR} codes.
Transaction times were lowest for audio and considerably higher for \gls{BLE}. \gls{QR} codes and audio had significantly higher usability than \gls{BLE}, highlighting the promising nature of the acoustic \gls{OOB} channel.
In our work, we build on these previous works to investigate the acoustic channel's suitability and usability for connecting multiple users simultaneously.

\subsection{Authentication Ceremonies in Secure Messaging Applications}
Modern \gls{E2EE} tools such as Signal or WhatsApp use opportunistic \gls{E2EE} (also known as \gls{TOFU}), requiring almost no user interaction. However, \gls{TOFU} cannot defend against active attacks unless users perform the \emph{authentication ceremony}, which is the process by which users can verify their cryptographic keys \cite{herzberg2021Secure,unger2015SoK}.
Authentication ceremonies usually require a human in the loop to match specific representations of the cryptographic fingerprint, such as hexadecimal strings, word phrases, or visualizations \cite{tan2017Can,dechand2016Empirical}.

The authentication ceremonies in \gls{E2EE} tools are not designed for multiple people and would require multiple pairwise instantiations, which is a prohibitive amount of effort in a collaborative scenario.
In addition to that, previous user studies have shown that even for just two users, current authentication ceremonies are too hard to use \cite{assal2015What,abu-salma2017Obstacles} and error-prone \cite{shirvanian2015Security,shirvanian2017Pitfalls}, which means that, in practice, people almost never perform them \cite{vaziripour2017that,vaziripour2018Action,wu2019Something,oesch2022User,schroder2016When,fassl2023Why}.
This is especially alarming considering that the default mode without performing the authentication ceremony fails to meet users' security expectations; most users are unaware of this issue \cite{herzberg2016Can}.

\subsection{SafeSlinger}
The most promising approach to date in literature is the SafeSlinger protocol by
Farb et al. \cite{farb2013SafeSlinger},
enabling the secure exchange of contact information between multiple people.
Built upon the recommendations of a previous user study on group pairing protocols \cite{nithyanand2010Groupthink}, \SafeSlinger surpasses prior human-in-the-loop approaches \cite{linSPATESmallgroupPKIless2009,chenGAnGSGatherAuthenticate2008} in usability, efficiency, and security, making it the state of the art in group pairing protocols.
It was designed specifically for groups of people to address the security and usability problems of using pairwise protocols in such scenarios.
Participants perform the following steps while using SafeSlinger:
\begin{enumerate}
    \item \textbf{Initialization.} Each participant must select the group size. Next, each participant sees a number on their smartphone. They must coordinate to identify and enter the lowest number among all participants on their smartphone.
    \item \textbf{Verification.} Each participant gets displayed three word phrases. They must coordinate to identify and select the phrase common to all smartphones.
    \item \textbf{Finalization.} The protocol then securely exchanges their contact information. Finally, each participant verifies that they received correct contact information from the others.
\end{enumerate}

In their evaluation, SafeSlinger was significantly faster and more usable than the pairwise contact exchange method Bump,\footnote{Bump was discontinued in 2014.} as shown by a within-subjects user study ($N=24$).
However, each step of the SafeSlinger protocol requires all participants to perform multiple comparisons with all other group members.
This increases effort for larger group sizes and is error-prone \cite{kainda2009Usability,tan2017Can}.
Furthermore, SafeSlinger requires a stable Internet connection during the pairing process and relies on a third-party server to facilitate the contact exchange.
These limitations prevent SafeSlinger from meeting our usability and deployability requirements (\autoref{sec:problem}).

\section{Research Questions}\label{sec:research-questions}\label{sec:research-hypothesis}

Based on our analysis of related work in the previous section, the main problem of previous approaches is the lack of usability for ordinary users, which gets amplified for larger groups of people due to inadequate scalability.
SafeSlinger is the most promising approach so far, but lacks in scalability and deployability.
The authentication ceremonies in \gls{E2EE} tools, as well as most current secure device pairing methods, are solely designed for two devices. 
Users expect a single associating interaction for the entire group \cite{chong2013How,jokela2014FlexiGroups,jokela2015Connecting}, but pairwise schemes cannot be easily generalized to multiple participants \cite{farb2013SafeSlinger}.

In this work, we explore the problem of secure contact exchange from a new angle, under the following research hypothesis: \textit{users prefer a contact exchange method with minimal user interaction.} 
Consequently, we design a new protocol named \OurApproach that minimizes user interaction, implement it on smartphones, and conduct a lab study comparing it with the current state-of-the-art approach, SafeSlinger.
The research questions we address in our work are:

\begin{itemize}
\item\textbf{RQ1.}
\emph{Which initialization and verification steps do users prefer?}

\item\textbf{RQ2.}
\emph{Which method has better usability?}

\item\textbf{RQ3.}
\emph{How do users perceive the security of both methods?}

\item\textbf{RQ4.}
\emph{How do users like the audible acoustic \gls{OOB} channel for pairing?}
% \medskip
\end{itemize}

\section{\OurApproach: Acoustic Group Pairing Protocol}\label{sec:design}

This section describes the design and implementation of our novel acoustic group pairing protocol, called \textit{\OurApproach}.
We developed it to meet the usability, deployability, and security requirements outlined in \autoref{sec:problem}.
Our starting point was the design of SafeSlinger,
which we identified as the most promising approach to date in \autoref{sec:related-work}, as it fulfills all our security requirements -- but not all usability and deployability requirements.

\begin{figure}
\begin{subfigure}[t]{0.25\columnwidth}
    \centering
		\includegraphics[width=\columnwidth]{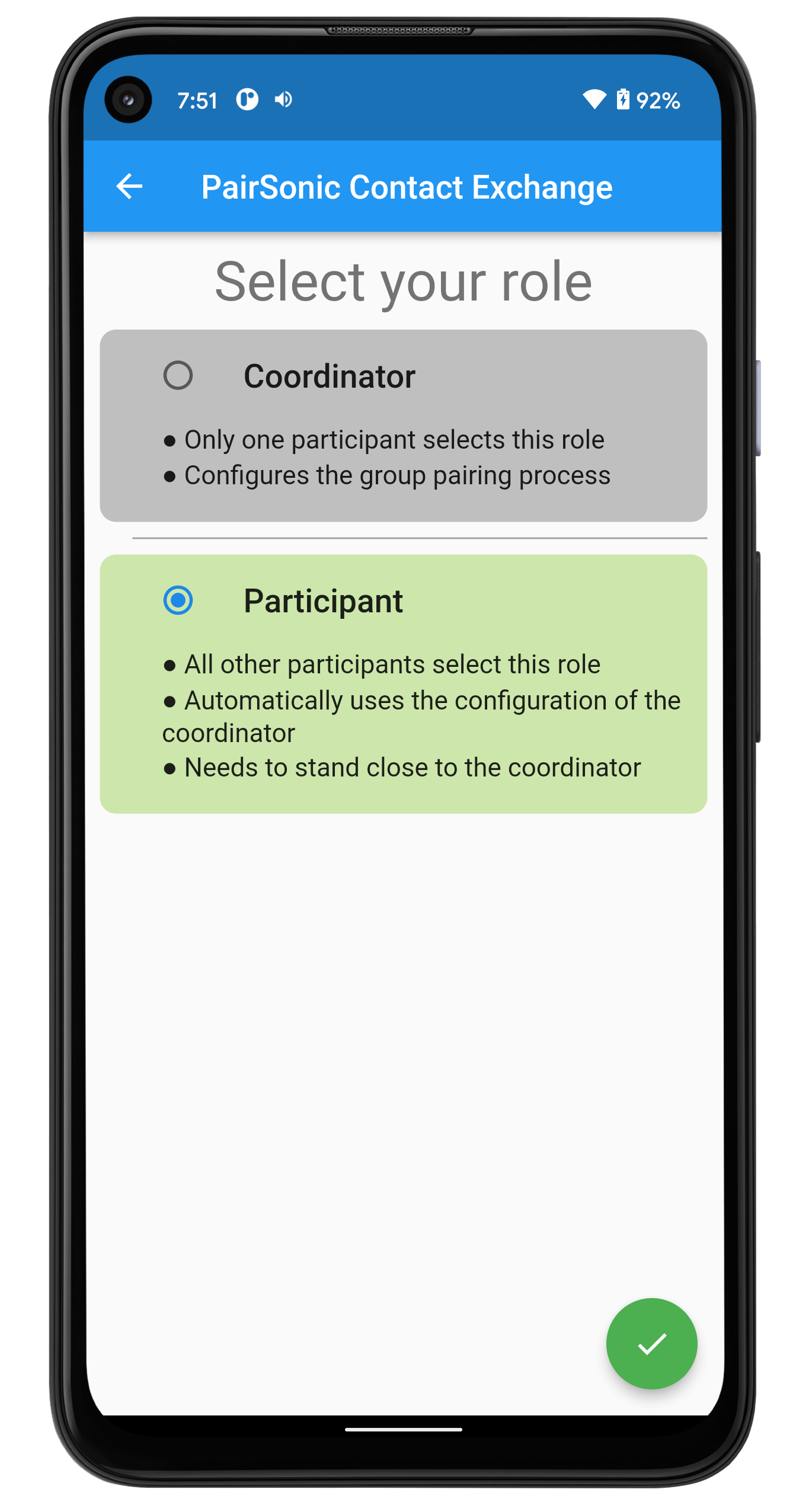}
    \caption
    {
        Initialization: Role.
    }
    \Description{%
        Role selection with a user interface on a smartphone screen. The user can select either the "Coordinator" or the "Participant" role.
	}
    \label{fig:pairsonic-participant1}
\end{subfigure}%
\begin{subfigure}[t]{0.25\columnwidth}
    \centering
		\includegraphics[width=\columnwidth]{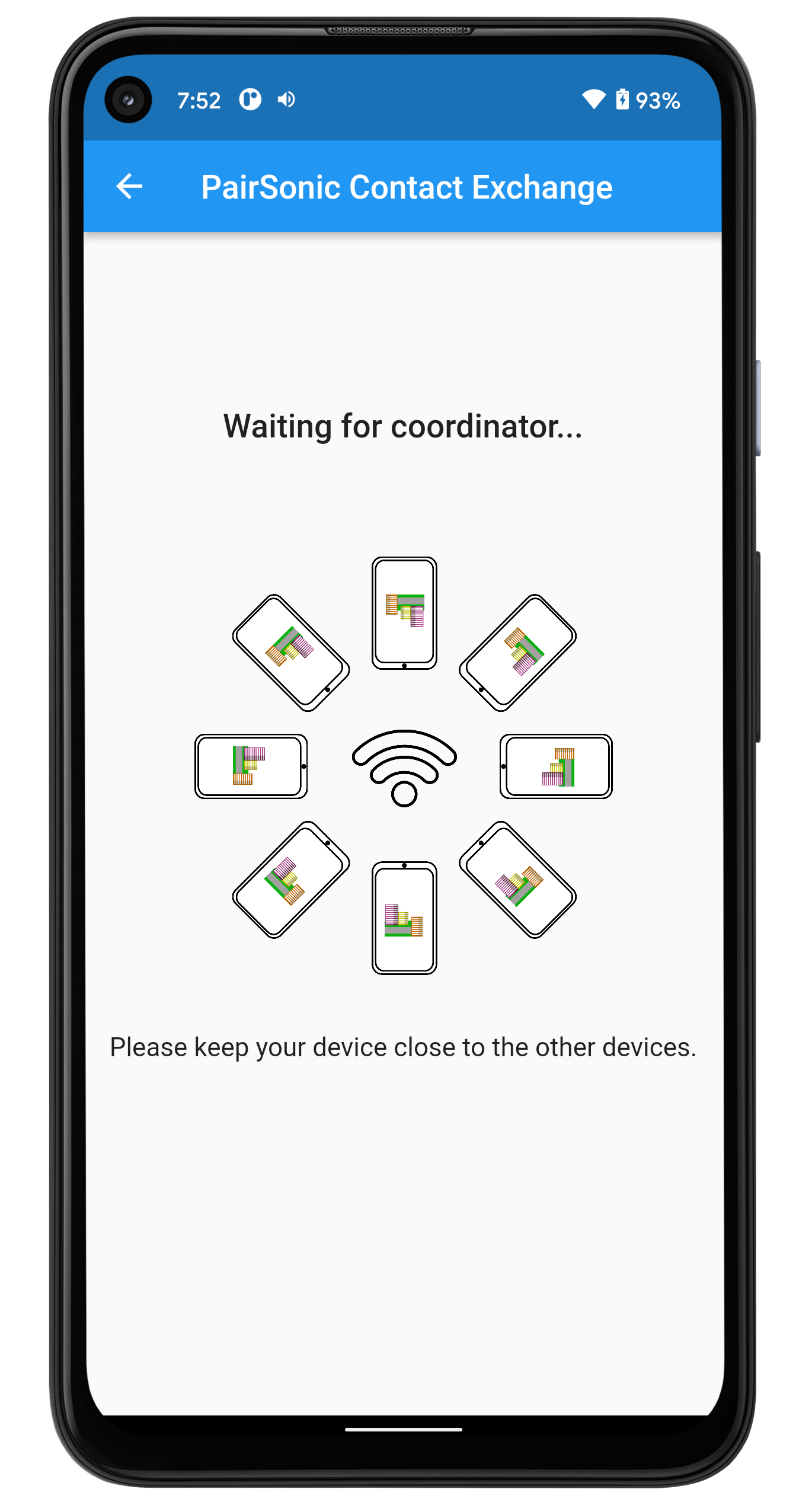}
    \caption
    {
        Initialization: Ready.
    }
    \Description{%
        Participants' phones displaying a "ready" status, indicating that they are now waiting for the coordinator.
	}
    \label{fig:pairsonic-participant2}
\end{subfigure}%
\begin{subfigure}[t]{0.25\columnwidth}
    \centering
		\includegraphics[width=\columnwidth]{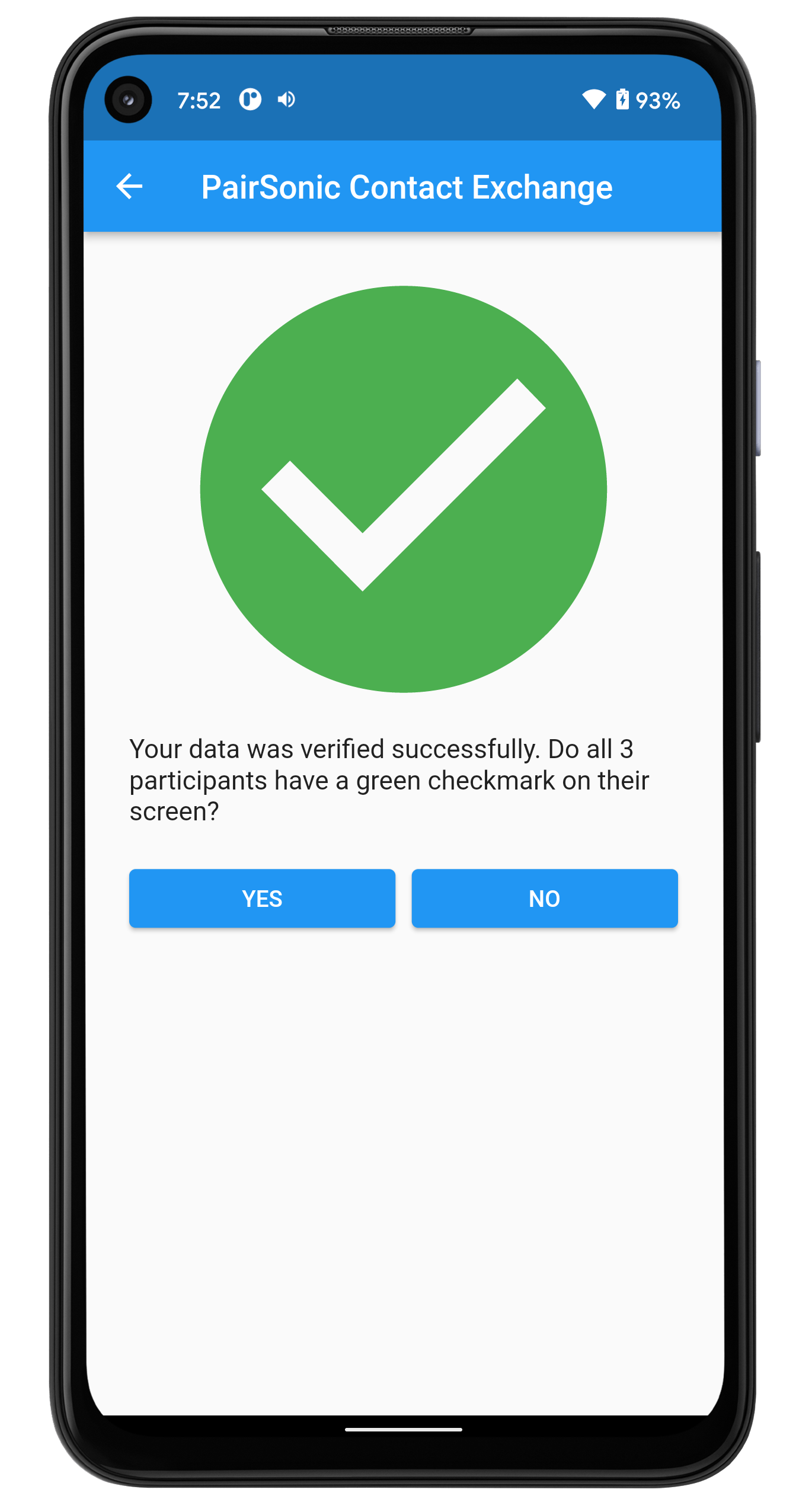}
    \caption
    {
        Verification.
    }
    \Description{%
        Verification phase with phones showing a green checkmark.
	}
    \label{fig:pairsonic-participant3}
\end{subfigure}%
\begin{subfigure}[t]{0.25\columnwidth}
    \centering
		\includegraphics[width=\columnwidth]{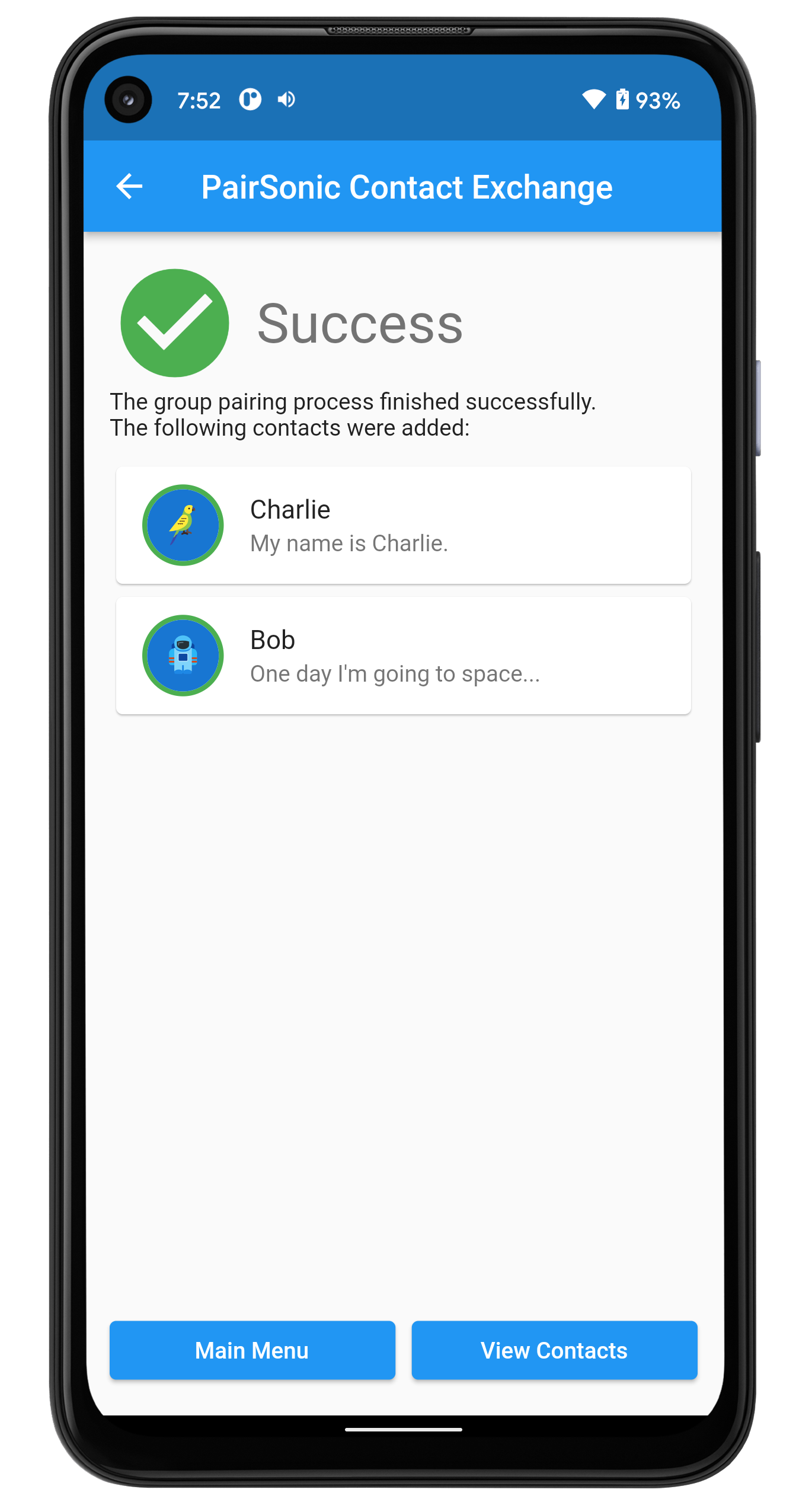}
    \caption
    {
        Finalization.
    }
    \Description{%
        Final screen displaying exchanged contact details.
	}
    \label{fig:pairsonic-participant4}
\end{subfigure}
\caption%
{
This figure shows the \OurApproach contact exchange from the perspective of the \textit{participant} role.
(a) The participants begin by selecting their role.
(b) They then wait for the coordinator to initiate the exchange via the acoustic \gls{OOB} channel, which is shown in \autoref{fig:ourapproach-smartphone-flow-coordinator}.
(c) Next, they verify that each group member's screen displays a green checkmark.
(d) After this process, the app shows the exchanged contacts.
}
\label{fig:ourapproach-smartphone-flow}
\Description{%
    A four-part composite image showing screenshots illustrating the steps of the PairSonic contact exchange process from the participant's perspective.
}
\end{figure}

\subsection{Protocol Overview}
\OurApproach, our innovative acoustic group pairing protocol, involves several steps as shown in \autoref{fig:ourapproach-smartphone-flow}.
First, a group of users wishing to exchange contact data initiates \OurApproach on their smartphones.
This brings them to the first screen, where they nominate a temporary \textit{coordinator} (\autoref{fig:pairsonic-participant1}).
The rest of the group assumes the role of \textit{participants}, which brings their devices into the ready state (\autoref{fig:pairsonic-participant2}).
The coordinator enters the total group size and confirms that all participants' smartphones are ready and nearby.\footnote{The coordinator's perspective of the \OurApproach contact exchange is depicted in \autoref{fig:ourapproach-smartphone-flow-coordinator} (appendix).}

Following this, the coordinator's smartphone creates a temporary ad-hoc WiFi network, transmitting a short acoustic message containing the network details to other smartphones. This allows them to automatically join the network.\footnote{\OurApproach can function even if a user is already connected to a WiFi network; the user will temporarily switch to the ad-hoc network instead.}

Next, the devices use this WiFi network to perform a cryptographic protocol, which is based on the \SafeSlinger protocol. Each participant's smartphone sends a nested commitment with their contact data\footnote{The exchanged data includes the cryptographic public key and, optionally, further information such as a profile picture.} to the coordinator's smartphone, which then disperses it to the other devices. 
The coordinator's smartphone generates a hash value from all participants' commitments and contact information, including their cryptographic public keys, and transmits this hash value over the acoustic \gls{OOB} channel.
Other smartphones are monitoring this channel and will abort if they receive any other message. Upon receipt of the correct hash value, a green checkmark appears on the smartphones, prompting users to confirm that all other smartphones display the same checkmark (\autoref{fig:pairsonic-participant3}).
Upon confirmation, the smartphones use the WiFi network to publish success nonces.
These allow others to verify the commitments and, if verification is successful, display the exchanged contacts (\autoref{fig:pairsonic-participant4}).
Conversely, if the checkmark is not confirmed, they release abort nonces, terminating the protocol for all participants. This procedure ensures that all participants receive accurate and verified contact information, including authenticated cryptographic public keys.

\subsection{Automating the User Interaction With Acoustic Communication}

To reduce user effort during the protocol, we automated the majority of \SafeSlinger's user interaction.
Instead of relying on users to act as data channels via error-prone shared device interactions, we utilize an \gls{OOB} channel to verify the integrity of the exchanged contact information by directly connecting the devices, thereby minimizing the required user interaction.
We use an acoustic location-limited channel where smartphones repurpose their integrated speakers and microphones to exchange data within a short radius of about one meter.
The smartphone speaker, rather than playing music, emits sound waves that encode information by varying their frequency or amplitude.
This process has been used to connect two devices in related work (\autoref{sec:related-acousticcommunication}), and we extend this concept to multiple devices.
The other smartphones capture the sound waves, decode the embedded data, and use it as an additional security layer to verify the integrity of the exchanged contact information.

The acoustic channel uniquely meets all our deployability requirements, with the added advantage that its physical layer is entirely customizable in software.
This means the security of the exchanged data can be directly protected on the physical layer \cite{putz2020AcousticIntegrityCodes}, an advantage when implementing our ad-hoc requirement without having to rely on existing security contexts such as pre-shared keys.

Another reason for choosing an acoustic \gls{OOB} channel is that the acoustic data transmission only works when the users are in proximity and bring their devices close to each other, as depicted in \autoref{fig:teaser}.
The physical interaction of moving all devices close to each other matches the intuitive behavior that users would naturally like to perform to connect their devices \cite{kray2010Userdefined,chong2011How}, similar to a handshake \cite{kuo2008Mind}.
In contrast, WiFi or Bluetooth also work over longer distances and even through walls, failing to ensure proximity.

Furthermore, we use the acoustic \gls{OOB} channel not only for verification but also to initiate the protocol, automatically connecting all participants' smartphones to the coordinator's WiFi Direct network.
Both WiFi and Bluetooth require a manual pairing process before communication can occur. We automate this process using acoustic communication, making the temporary network discovery and association as part of our group pairing protocol effortless and seamless.

\subsection{\OurApproach Implementation}\label{sec:implementation}
We implemented \OurApproach as an Android \gls{app} using the Flutter framework and utilized Android's WiFi Direct API for local ad-hoc WiFi functionality.
The acoustic \gls{OOB} channel was built using the ggwave library.\footnote{Data-over-sound library ggwave: \url{https://github.com/ggerganov/ggwave}}
The physical layer uses multi-frequency \gls{FSK} modulation with 96 evenly spaced frequencies ranging from \SI{1875}{\hertz} to \SI{6375}{\hertz}, according to ggwave's \texttt{AUDIBLE\_FAST} profile.
We selected an audible frequency range to assess user perception of this form of communication.
Practical applications could choose between audible or inaudible frequency ranges, based on factors such as robustness, usability, and hardware compatibility.

As our implementation is a custom prototype using a novel communication stack combining acoustic communication with WiFi Direct,
we conducted functionality and compatibility tests with smartphones from various manufacturers before our lab study.
Our prototype worked on all tested devices supporting WiFi Direct, with a minimum required Android version 6.0 (API level 23, released in 2015): LG Nexus 5X, Huawei Nexus 6P, Google Pixel 4, Pixel 4a, Pixel 5, Pixel 6 Pro, Samsung Galaxy S20 Ultra, Galaxy S22, Oppo Reno 6, OnePlus 10 Pro, Xiaomi 11T Pro.
Our design is principally also compatible with iOS, given the ubiquitous compatibility of acoustic communication.
However, since iOS does not support WiFi Direct, this channel would need to be replaced with another communication layer supported on Apple devices, such as \gls{BLE} or AWDL \cite{stute2018awdl}. Our design does not inherently depend on any specific radio channel, making such a modification feasible.

\subsection{Comparison to \SafeSlinger}
Our design's key differentiator is the automation of as many manual steps as possible, resulting in a group pairing protocol that is effortless for the user.
Such a design allows us to verify our research hypothesis by comparing SafeSlinger with \OurApproach in a user study.
\autoref{tab:protocol-comparison} lists the main differences between \OurApproach and SafeSlinger, which we now discuss in more detail.

\begin{table}[t]
\renewcommand{\arraystretch}{1.1}
\centering
\caption{Overview of the two group pairing protocols compared in our study.}
\tablefontsize
\begin{tabularx}{\columnwidth}{@{} l XX @{}}
\toprule
&
\tableheadline{\SafeSlinger} & 
\tableheadline{\OurApproach} \\
\midrule
\tableheadline{User Grouping} &
\begin{tableitemize}
    \item peer-based
\end{tableitemize} & 
\begin{tableitemize}
    \item leader-based
\end{tableitemize} \\
%\midrule
\rowcolor{\altrowcolor}
\tableheadline{Device Communication} & \begin{tableitemize}
    \item Internet via central server
\end{tableitemize} &
\begin{tableitemize}
    \item decentral via WiFi ad-hoc
    \item local acoustic communication
\end{tableitemize} \\

%\midrule
\tableheadline{Phase 1: Initialization} & \begin{tableitemize}
    \item each participant counts + enters group size
    \item each participant compares ID with others and enters lowest ID
\end{tableitemize} &
\begin{tableitemize}
    \item only leader counts + enters group size
    \item participants bring devices into proximity
\end{tableitemize} \\
%\midrule
\rowcolor{\altrowcolor}
\tableheadline{Phase 2: Verification} &
\begin{tableitemize}
    \item each participant compares three 3-word phrases with others and selects matching phrase
\end{tableitemize} &
\begin{tableitemize}    \item each participant confirms whether all devices show green checkmark
\end{tableitemize} \\
%\midrule
\tableheadline{Phase 3: Finalization} &
\begin{tableitemize}    \item each participant verifies and selects which contact entries to import
\end{tableitemize} &
\begin{tableitemize}    \item each participant verifies list of new contact entries
\end{tableitemize} \\
\bottomrule
\end{tabularx}
\label{tab:protocol-comparison}
\end{table}

\subsubsection{Usability}
Whereas in SafeSlinger all participants must enter the group size in the initialization phase, in \OurApproach only one participant needs to enter the group size.
This smartphone then starts broadcasting initialization information to the other participants' smartphones via the acoustic \gls{OOB} channel.
The other participants only need to bring their devices close together.
The second part of SafeSlinger's initialization phase,
which requires all participants to coordinate to find and enter the lowest number shown on their devices, is completely automated in \OurApproach using the acoustic \gls{OOB} channel.

After exchanging their contact information, the participants need to verify that no adversary interfered with the protocol.
Whereas SafeSlinger requires all participants to find a matching word phrase among three options, we automate this using an acoustic verification message in \OurApproach.
This reduces the information that participants need to verify manually to a single bit, which is represented by a clear green checkmark or a red warning symbol. The participants must abort the process in case not all devices show the green checkmark.
In \OurApproach, the devices automatically verify a second acoustic message sent by the initiating smartphone to determine whether it matches the data received over the WiFi channel.
\OurApproach minimizes the required user interaction compared to SafeSlinger, without reducing the security guarantees.

In the final phase of the protocol, the participants must verify that the exchanged contact entries correspond to the involved participants.
This step cannot be automated as it is a social decision to determine if a contact entry matches the person standing in front of them, rather than a technical one. In \SafeSlinger, each participant also has to select which contact entries to import. In \OurApproach, we defer this decision by importing all contact entries, considering that users can manually delete contact entries whenever they want.

\subsubsection{Deployability}\label{sec:design-comparison-deployability}
In addition to usability improvements, our design also addresses the deployability limitations of \SafeSlinger by decentralizing the protocol.
Whereas \SafeSlinger requires a stable Internet connection during the pairing process, \OurApproach operates completely offline.
Thus, it can function even in areas without WiFi or cellular connectivity, such as rural zones, emergency situations, or developing countries without Internet access.
Our protocol does not require a third-party server because it operates in a decentralized manner, i.\,e., directly between the participants. This results in greater availability and resilience since our pairing process does not depend on external infrastructure.
It also offers a privacy advantage, as no metadata reaches any third party.

\subsubsection{Security}
While \OurApproach fundamentally changes both the user interaction and physical layer of SafeSlinger, it retains the same cryptographic protocol, thereby preserving the strong security benefits from SafeSlinger.
\OurApproach thus fulfills all security requirements of confidentiality, contact authentication, and collective pairwise security.
We rely on SafeSlinger's cryptographic primitives, namely hierarchical multi-value commitments and Group Diffie-Hellman key agreement (for further information, see Farb et al. \cite{farb2013SafeSlinger}).
Additionally, we employ an acoustic \gls{OOB} channel for location-limited verification as an extra security layer.

\section{Study Methods}\label{sec:methods}
To answer our research questions on the usability and security of the group pairing protocols \SafeSlinger and \OurApproach, we conducted a comparative lab study featuring practical contact exchange tasks (\autoref{sec:methods-material}), followed by a quantitative questionnaire and qualitative interviews (\autoref{sec:methods-design}).
We also discuss our study sample (\autoref{sec:methods-participants}), ethical considerations (\autoref{sec:methods-ethics}), and our analysis methodology (\autoref{sec:methods-analysis}).

\subsection{Material}\label{sec:methods-material}
Our goal for the study setting was to resemble a typical contact exchange scenario.
We built a dummy smartphone \gls{app} of a fictional social networking platform consisting of user profiles, direct messaging functionality, and public message boards.
This \gls{app} was built using Flutter targeting Android smartphones and supports \gls{E2EE} messaging using public-key cryptography.
Each user has a profile consisting of a username and additional profile information such as an avatar and a profile description. Additionally, the \gls{app} automatically generates public-key credentials for each user, which is used to secure the communication.
The app's main menu has a button to initiate the contact exchange process, triggering either \SafeSlinger or \OurApproach.

\begin{figure}
\begin{subfigure}[t]{0.25\columnwidth}
    \centering
		\includegraphics[width=\columnwidth]{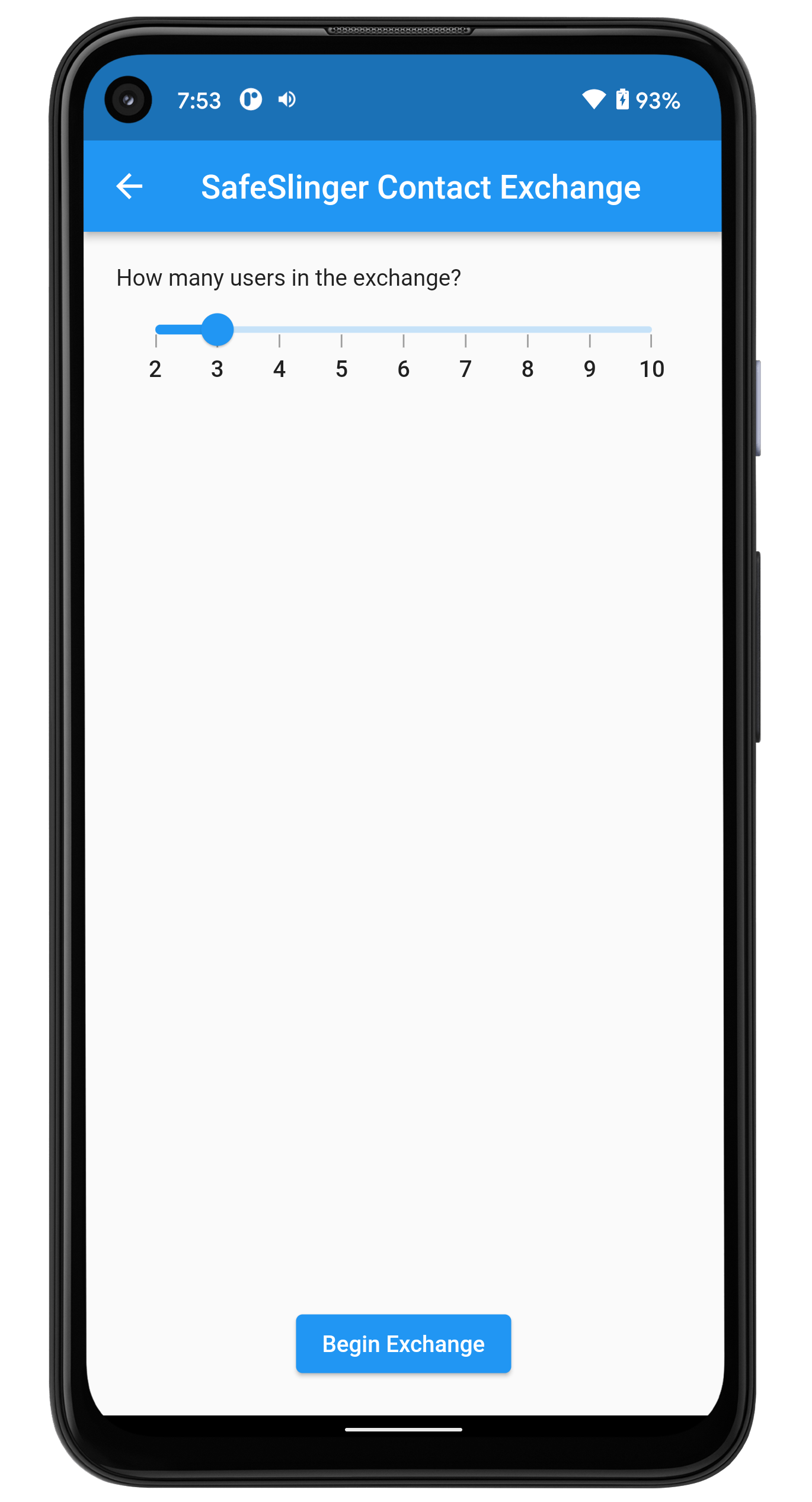}
    \caption
    {
        Initialization: Count.
    }
    \Description{%
        Users input the group size into the app using a slider interface.
	}
    \label{fig:safeslinger1}
\end{subfigure}%
\begin{subfigure}[t]{0.25\columnwidth}
    \centering
		\includegraphics[width=\columnwidth]{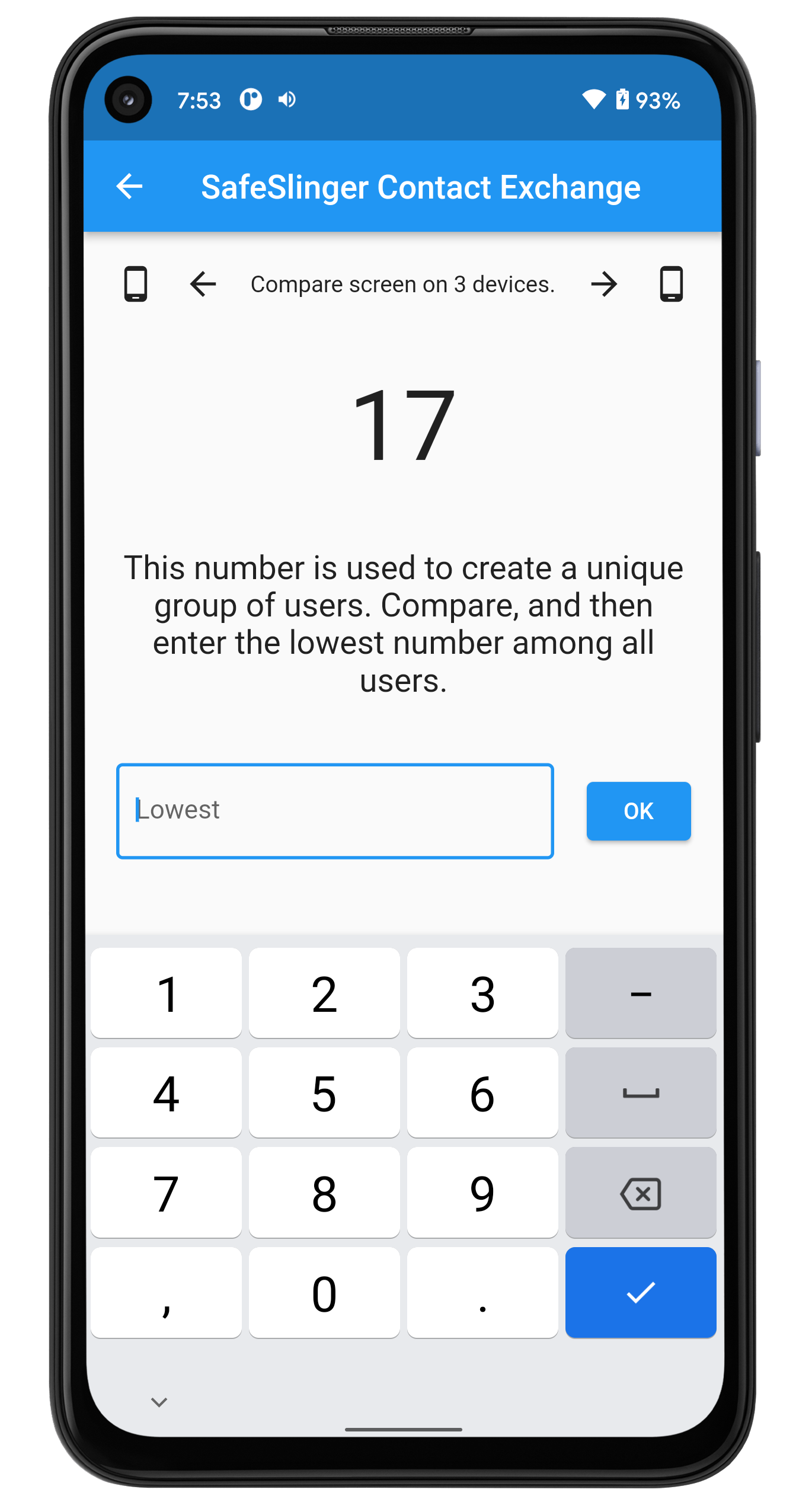}
    \caption
    {
        Initialization: Lowest ID.
    }
    \Description{%
        Participants are prompted to enter the smallest ID among the group, as shown by the number 17 on the screen.
	}
    \label{fig:safeslinger2}
\end{subfigure}%
\begin{subfigure}[t]{0.25\columnwidth}
    \centering
		\includegraphics[width=\columnwidth]{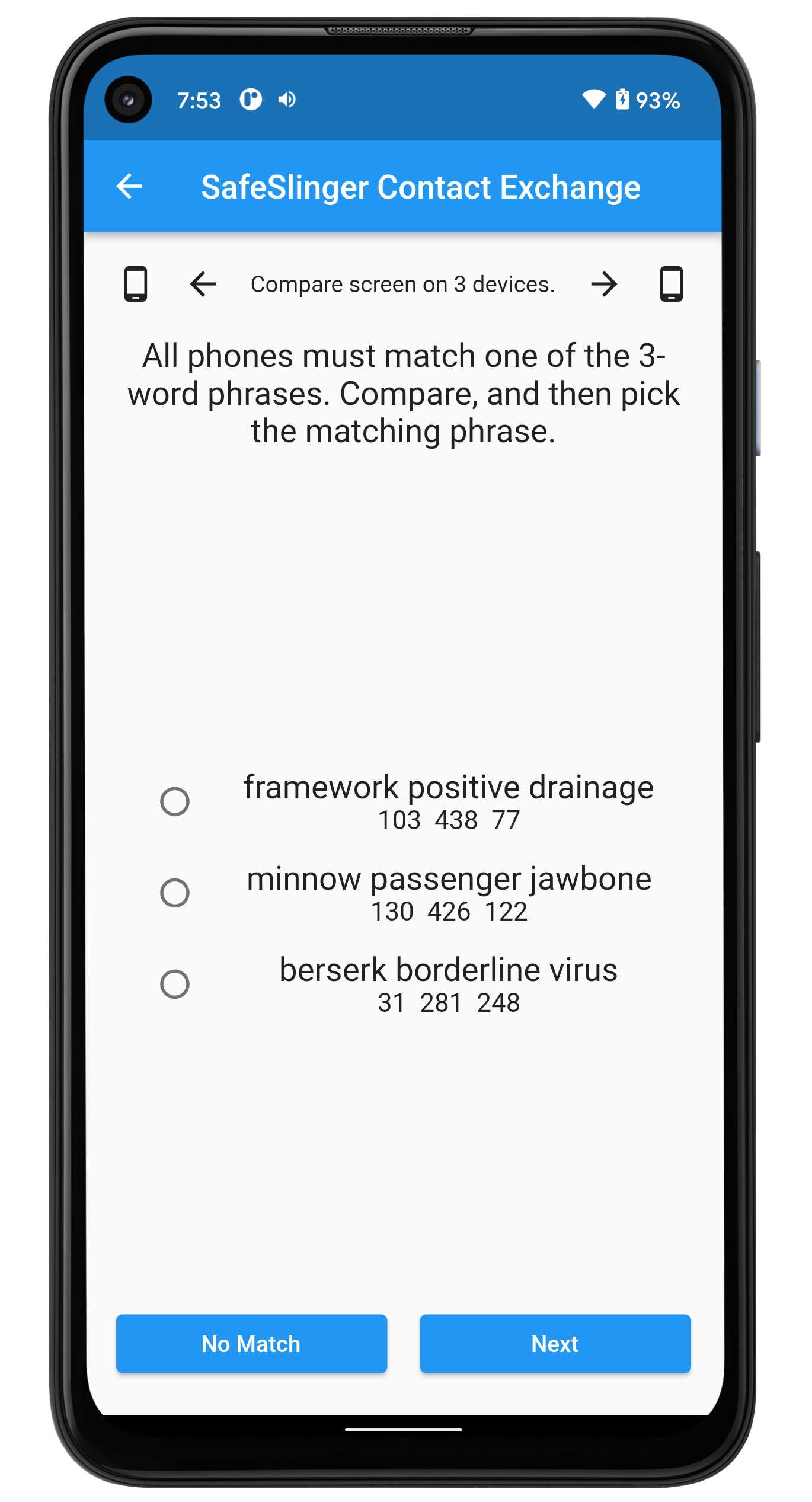}
    \caption
    {
        Verification.
    }
    \Description{%
        Users must match one of the three-word phrases shown on all phones for verification. This is done by selecting one of three options.
	}
    \label{fig:safeslinger3}
\end{subfigure}%
\begin{subfigure}[t]{0.25\columnwidth}
    \centering
		\includegraphics[width=\columnwidth]{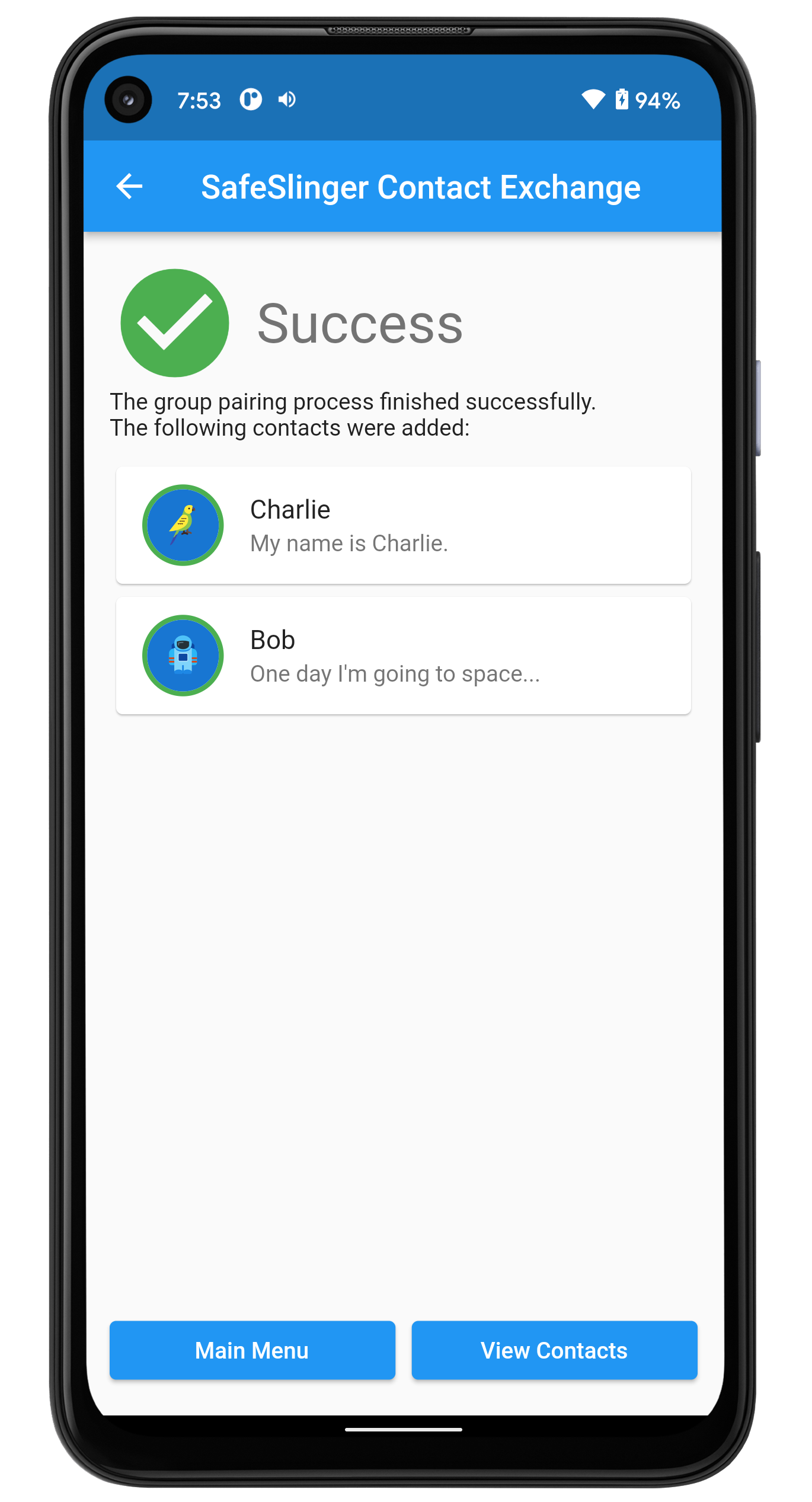}
    \caption
    {
        Finalization.
    }
    \Description{%
        The finalization screen confirms success, displaying exchanged contact details like usernames and avatars.
	}
    \label{fig:safeslinger4}
\end{subfigure}
\caption%
{
This figure shows the \SafeSlinger contact exchange, which we integrated into our app using the official Android library, only adjusting its aesthetics to align with our app's design.
(a) The \SafeSlinger contact exchange process begins with each participant selecting the total number of group members. (b) Next, all members compare their IDs to identify and input the smallest ID. (c) During the subsequent verification phase, members compare their word phrases to find and choose the phrase that appears on all devices. (d) After this process, the app displays the exchanged contacts.
}
\label{fig:safeslinger-smartphone-flow}
\Description{%
    A four-part composite image showing screenshots of the steps in the SafeSlinger contact exchange process integrated into an app. Steps include counting group members, selecting the lowest ID, matching phrases for verification, and displaying exchanged contacts.
}
\end{figure}

We integrated \SafeSlinger into our \gls{app} using the official Android library\footnote{SafeSlinger Android client library: \url{https://github.com/SafeSlingerProject/exchange-android}} by Farb et al. \cite{farb2013SafeSlinger}, including their user interface. We only modified the look and feel to match the rest of our \gls{app}. \autoref{fig:safeslinger-smartphone-flow} shows the different steps that participants have to perform using \SafeSlinger.
We used Google's cloud platform AppEngine to host SafeSlinger's server component\footnote{SafeSlinger Python server: \url{https://github.com/SafeSlingerProject/SafeSlinger-AppEngine}} written in Python.

\subsection{Study Design}\label{sec:methods-design}

We followed a mixed methods approach, combining usability testing within a lab study, open-ended oral interviews, and a post-test questionnaire.
We performed our lab study with different groups of two to six participants, who got to exchange their contact information amongst each other in an actual group scenario.
We used a within-subjects design to identify the differences between \OurApproach and the state of the art (\SafeSlinger{}), meaning that each participant performed two rounds of tasks within the same group of participants to gain hands-on experience with both systems.
We used counterbalancing to vary the order in which the participants encountered these systems, which reduced practice and boredom effects.
We decided for a within-subjects design due to greater statistical power than a between-subjects design.
We successfully tested our study setup using a pilot study with six participants.

We conducted our lab study in a neutral meeting room, where each group of participants met the study conductor, who was always present during each trial to verify that the participants performed the tasks and to conduct the interviews.
The lab setting in a neutral meeting room with the same smartphones\footnote{We always used the same smartphone models for the experiment, in the following order depending on the group size: 3$\times$Google Pixel 4a, Google Pixel 5, Samsung Galaxy S22, Oppo Reno 6.}
ensured consistent conditions for all participants, minimizing confounding factors.
During the study, the participants filled in our questionnaire section by section after each experiment round.
The study conductor additionally interviewed them about their impressions after the experiment.

\subsubsection{Briefing}
Participants first reviewed and signed our consent form, which outlined the study's aim, procedure, and privacy policy.
We then explained the study's agenda, briefly introducing the problem of securely exchanging contact information and how current approaches fail for larger groups (approximately four minutes total).
We used the same slide set for all groups to increase the internal validity of our study by ensuring that all participants received the same information.
The goal of the briefing was to ensure that all participants had the same understanding of the systems they were about to try, reducing potential confounding effects from misunderstandings.
Importantly, we did not disclose that we had designed one of the systems ourselves.

We then handed each participant the questionnaire (\autoref{sec:questionnaire}) and a smartphone, and asked them to fill in their profile information within the \gls{app} (name, avatar from a gallery of options, freely chosen text description; shown in \autoref{fig:app-profile}).
We gave each participant a choice between using their real name or a pseudonym within the \gls{app}.

\subsubsection{Experiment}\label{sec:methods-experiment}
Upon distributing the smartphones, we conducted our experiment in two rounds (A and B) corresponding to both group pairing systems. Each round was structured identically, but we alternated the order of \SafeSlinger and \OurApproach between each group to neutralize order effects.

\begin{figure}
\begin{subfigure}[t]{0.25\columnwidth}
    \centering
		\includegraphics[width=\columnwidth]{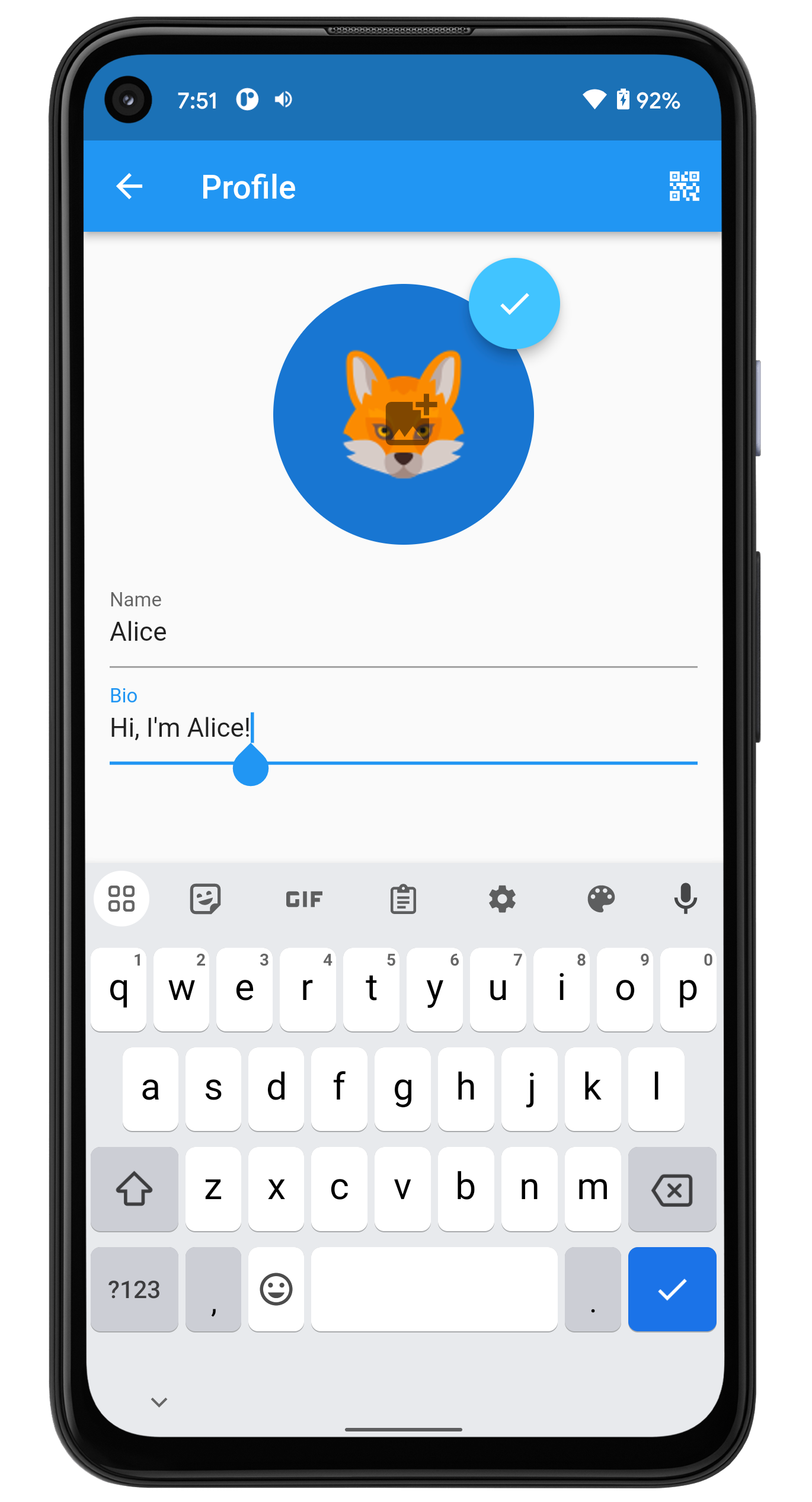}
    \caption
    {
    Profile screen.
    }
    \Description{%
        A smartphone screen displaying a user profile setup with a profile picture of a cartoon fox, username "Alice", and text entry for a bio.
	}
    \label{fig:app-profile}
\end{subfigure}%
\begin{subfigure}[t]{0.75\columnwidth}
    \centering
		\includegraphics[width=0.86\textwidth]{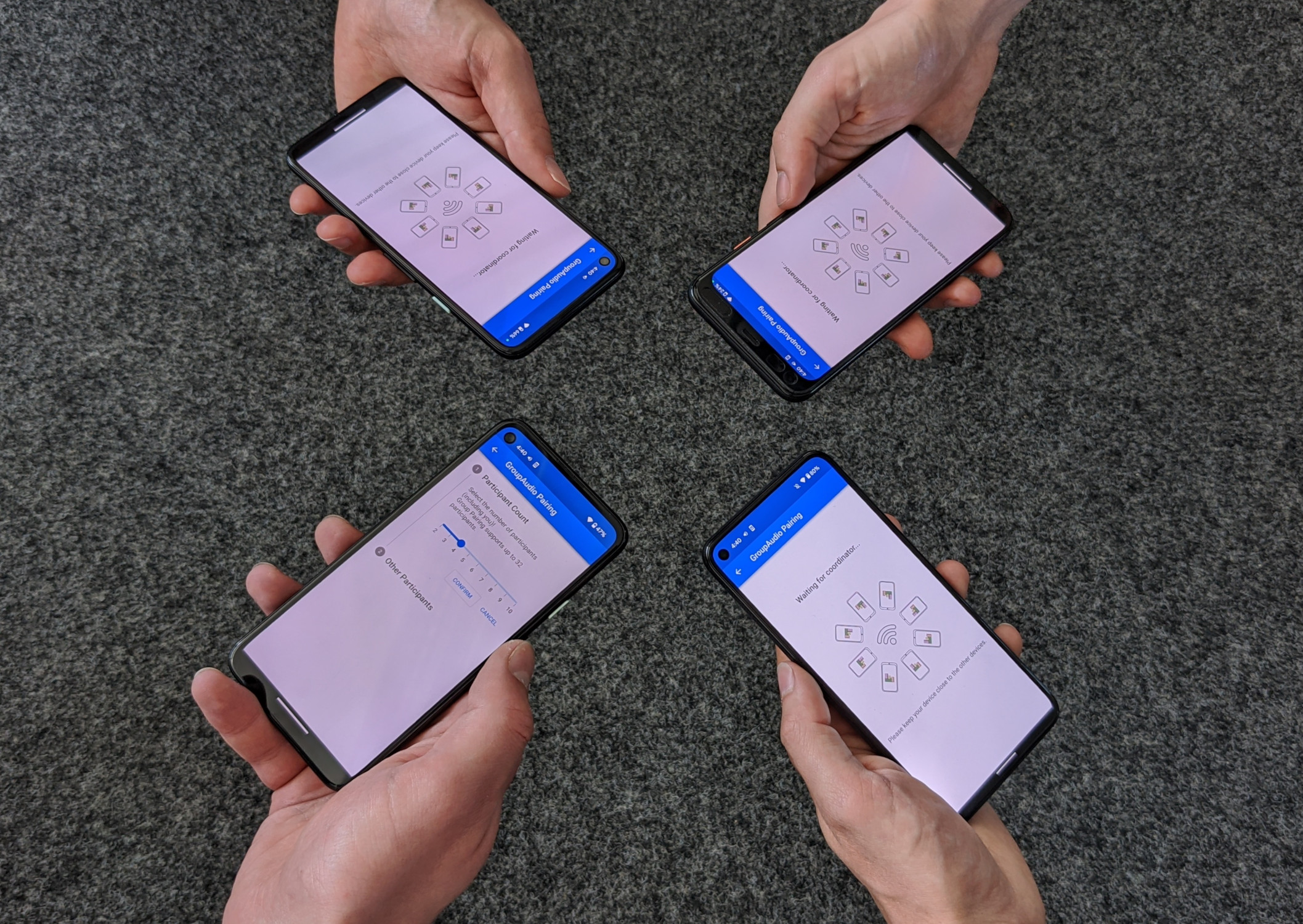}
    \caption
    {
    User group exchanging their contact information using \OurApproach.
    }
    \Description{%
        Four individual hands in a circle, holding their smartphones with the screens facing each other, in the process of exchanging contact information via PairSonic, indicated showing the pairing screen.
	}
    \label{fig:users-with-smartphones}
\end{subfigure}
\caption%
{
Our participants first created their profiles in the app, then collaboratively tested both contact exchange systems with the other group members. During \OurApproach, the participants bring their devices close together.
}
\Description{%
    A two-part image with a screenshot and a photo showing the PairSonic pairing process conducted in the study.
}
\end{figure}

Each round started with a warm-up task, where we asked the group to try the contact exchange system on their own without further guidance except the prompts within the \gls{app}.
\autoref{fig:users-with-smartphones} illustrates how \OurApproach's context exchange looked like for a group of four users.
The study conductor noted down potential problems in the study protocol and clarified the process with the group after they successfully exchanged their contact information.

We then asked the group to use the system a second time, but remotely and unbeknownst to the participants we manipulated the experiment in such a way as to simulate an active attack on the pairing process.
This resulted in the protocol failing on one of the smartphones.
The study conductor observed the group reacting to this situation and asked them why they thought the process failed.
We opted to simulate an attack because this provides a controllable and reproducible method to ensure all participants have a consistent experience.
By simulating rather than actively disrupting the wireless communication, we also avoided technical risks to nearby networks.
Our simulated attack mirrors the maximum capabilities of our theoretical adversary, who can remotely control smartphone radio communication undetected (\autoref{sec:adversary-model}). Incorporating this attack into our study helped us evaluate the usability of \OurApproach, especially in adversarial scenarios our system aims to counter.

After explaining that this protocol failure could indicate a security problem, we asked the group to use the system a third time.
We remotely disabled the attack simulation and told them that this time there would be no attack.
At the same time, we measured the time it took the group to perform this third task from the beginning till the end, to get an assessment of how fast the system works after some practice.

After the first round, we collected all smartphones, switched to the other group pairing protocol, and conducted the second round likewise.
Before concluding the study, we thanked the participants for their time. In total, each of our participants spent 60 minutes to take part in our study.

\subsubsection{Questionnaire}\label{sec:methods-questionnaire}
After each of the two rounds, participants filled in the respective sections of the questionnaire to reflect on their experiences during the experiment tasks. Once both rounds were completed, they filled in sections about their preference of both system, about their usage patterns of different social or collaborative tools, and demographic questions at the end.
Our corresponding questionnaire, presented in \autoref{sec:questionnaire}, was used to collect the following variables:

\begin{itemize}
    \item \textbf{Dependent variables.}
    The dependent variables captured our participants post-test assessment of both \OurApproach and \SafeSlinger. For each system, we used the \gls{SUS} \cite{brooke1996sus} to answer RQ2 on usability. Because efficiency is one aspect of usability \cite{rubin2008Handbook}, the study conductor also measured the time it took the groups to complete the protocol during the third task. To answer RQ3, we asked an additional question about the perceived security of the systems, using a five-level Likert item.
    \item \textbf{Preference.} To answer RQ1, this binary variable captures the participants' preference: either they preferred \OurApproach or \SafeSlinger.
    \item \textbf{Collaborative/social tool usage.} To measure our participants' use of collaborative platforms and tools, we provided a list of nine distinct types of digital communication groups. For each type, we asked the participants to state whether they actively took part in them, the average number of participants in such groups, and if security was important to them in these contexts. Participants could also provide information about additional group types using free text fields. Additionally, we requested them to estimate the total number of digital communication groups they overall participate in.
    \item \textbf{Control variables.}
    We used two control variables: participants' prior experience with smartphones and their \gls{ATI} \cite{franke2019affinity}. Both variables help to assess the background of our sample and to explore potential correlations between technology affinity and the usability or security of both systems \cite{chong2013How}.
    As a standardized measure, the \gls{ATI} also improves the utility of our dataset for future research on usable security.
    \item \textbf{Demographic variables.} We collected the gender, age, education, field of study/work, and student status of our participants as demographic variables.
\end{itemize}

\subsubsection{Interview}
At four stages during the study, we interviewed our participants with open-ended qualitative questions to better understand their behavior and reasoning. All participants agreed to be recorded in our consent form. The study conductor also asked follow-up questions to learn more about the intricacies of their reasoning, but always asked the same set of starting questions for each group, in accordance with our research questions.

\medskip\noindent
After filling in each round's questionnaire section about the prior used system, we asked participants:
\begin{itemize}
    \item What did you like or dislike about this system?
    \item How secure do you think this system is?
\end{itemize}

\medskip\noindent
Once participants gave feedback to each approach individually and completed the questionnaire about their general preference, we interviewed participants to evaluate the two approaches in direct comparison:
\begin{itemize}
    \item Which system do you prefer, and why?
    \item \textit{$\langle$Study conductor clarifies which part of each system constitutes the initialization step.$\rangle$} Which initialization step do you prefer, and why?
    \item \textit{$\langle$Study conductor clarifies which part of each system constitutes the verification step.$\rangle$} Which verification step do you prefer, and why?
\end{itemize}

\medskip\noindent
After the participants filled in the questionnaire section about their usage patterns for various types of collaborative tools, we asked about security:
\begin{itemize}
    \item How important is secure communication to you in group or collaborative scenarios?
\end{itemize}

\subsection{Participants}\label{sec:methods-participants}

\begin{table}
\caption{Our participants' demographic data and control variables. We report percentages (and frequencies).}
\label{tab:demographics}
\begin{minipage}[t]{.48\linewidth}
    \vspace{0pt}
    \centering
    \tablefontsize
    \begin{tabularx}{\columnwidth}{lXr}
    \toprule
    &
    \tableheadline{Variable} &
    \tableheadline{Our Sample}\\
    \midrule
    \multirow{4}*{Gender} & Female & 40\% (18) \\
    & Male & 49\% (22) \\
    & Diverse & 4\% (2) \\
    & No answer & 7\% (3) \\
    \midrule
    \multirow{4}*{Age} & 18--19 & 2\% (1) \\
    & 20--24 & 49\% (22) \\
    & 25--29 & 36\% (16) \\
    & 30--34 & 11\% (5) \\
    %& 35--39 & 0\% (0) \\
    & 40--44 & 2\% (1) \\
    \midrule
    \multirow{2}*{Student} & Yes & 82\% (37) \\
    & No & 18\% (8) \\
    \midrule
    \multirow{2}*{Smartphone Exp.} & $>$ 2 years & 98\% (44) \\
    & $\le$ 2 years & 2\% (1) \\
    \midrule
    \multirow{1}*{ATI} & Median & 4.6 \\
    & IQR & 1 \\
    \bottomrule
    \end{tabularx}
\end{minipage}\hfill
\begin{minipage}[t]{.48\linewidth}
    \vspace{0pt}
    \centering
    \tablefontsize
    \begin{tabularx}{\columnwidth}{lXr}
    \toprule
    &
    \tableheadline{Variable} &
    \tableheadline{Our Sample}\\
    \midrule
    \multirow{6}*{School Education} & Intermediary secondary & 4\% (2) \\ \addlinespace
    & University entrance qualification & 93\% (42) \\ \addlinespace
    & No answer & 2\% (1) \\
    \midrule
    \multirow{8}*{Prof. Education} & Vocational training & 11\% (5) \\ \addlinespace
    & Technical college & 4\% (2) \\ \addlinespace
    & Bachelor & 22\% (10) \\
    & Master & 22\% (10) \\
    & PhD & 2\% (1) \\ \addlinespace
    & No degree & 36\% (16) \\
    \bottomrule
    \end{tabularx}
\end{minipage} 
\end{table}

We conducted our study with 47 participants between January and March 2023. For recruitment, we used mailing lists, social media groups, word-of-mouth, and snowball sampling, both within and outside our university. Participants had to be over 18 years old to be eligible.
Interested participants self-registered on an online form to choose all suitable time slots for voluntary participation.
We randomly distributed our participants into groups of two to six participants based on their availability and invited them to our lab study.
This resulted in five groups with four people, two groups with each of two, three, and six people, and one group with five people, for a total of 12 unique groups.
Each participant was compensated 15 EUR (in cash) for taking part in our study.

Two participants did not fully complete the questionnaire, so we removed them from our final sample ($N = 45$).
Of these, 22 identified as male, 18 as female, 2 as diverse, and 3 preferred not to answer their gender.
While 16 participants did not have any professional or university degree, the education level of our participants was generally high, with 10 participants having a Master's degree. 
Most participants (37) were students.
Our participants were aged 18--44; most participants (38) were between 20 and 29 years old (\autoref{tab:demographics}).

\subsection{Ethical Concerns}\label{sec:methods-ethics}
Our university's \gls{IRB} reviewed and approved this study.
We gathered written consent from all participants after informing them about the study's purpose and data collection in compliance with the \gls{GDPR}.

We did not collect any sensitive data, and we made answering demographic questions in our questionnaire optional.
Participants had the voluntary choice to permit interview recordings; all agreed, so we recorded all interviews and deleted the files after transcription to safeguard participants' privacy.
We stored each participant's responses pseudonymously, using sequential numbers without including any identifying information.
Participants voluntarily engaged in our study and received compensation higher than our country's minimum wage as recognition for their time and effort.

\begin{table}
\centering
\caption{Comparison of the dependent variables.}
\tablefontsize
\begin{tabularx}{\columnwidth}{Xrrrr}
\toprule
\tableheadline{Dependent Variable} &
\tableheadline{\SafeSlinger} &
\tableheadline{\OurApproach} &
\tableheadline{Statistic} &
\tableheadline{ES}\\
& ($N=45$) & ($N=45$) \\
\midrule
System Usability Scale (SUS) & 75 & 85 &$V = 597.5$&$r=-.43$\\
    & (25) & (17.5) & $\bm{p=.004}$&\\
\rowcolor{\altrowcolor} Security & 4 & 4 &$P = .5$&$g=0$\\
\rowcolor{\altrowcolor}    & (1) & (1) & $p=1$&\\
Preference & 31\% (14) & 69\% (31) &$P = .69$&$g=.19$\\
&  &  & $\bm{p=.016}$&\\
%\midrule
\rowcolor{\altrowcolor}Completion Time (Groupwise, $N=12$) & \SI{35.5}{\second} & \SI{33.5}{\second} & $V = 49.5$ & $r=-.23$ \\ % N=12; groupwise metric!
\rowcolor{\altrowcolor}& (\SI{8.5}{\second}) & (\SI{10}{\second}) & $p=.432$&\\
\bottomrule
\tablenotesx{5}{\textit{Note:} for the binary preference rating, we report percentages (and frequencies), the binomial test, and Cohen's $g$ \cite{Cohen1992effectsize} as the effect size (ES). For the other variables, we report the median (and IQR).
For the \gls{SUS} and completion times, we additionally report the Wilcoxon signed-rank test, and the ES estimate based on Rosenthal's method \cite{rosenthal1991meta};
for the security ratings, we report the sign test and Cohen's $g$.}
\end{tabularx}
\label{tab:descriptive-dependent}
\end{table}

\subsection{Data Analysis}\label{sec:methods-analysis}
We used R 4.2.2 for all quantitative data analysis \cite{r2023language} and MaxQDA 2022 for the qualitative data coding \cite{maxqda2022}.
We calculated the central tendencies and correlations to answer our research questions.
We compared \SafeSlinger and \OurApproach using the Wilcoxon signed-rank test \cite{wilcoxon1945individual} for interval variables, the sign test \cite{siegel1956nonparametric} for ordinal variables, Pearson's chi-squared test \cite{pearson1900criterion} for nominal variables, and the binomial test \cite{siegel1956nonparametric} for proportions.
We report descriptive statistics for our quantitative variables using the median (Mdn) and the interquartile range (IQR).
We used Kendall's rank correlation coefficient \cite{kendall1938measure} to determine the relationship between our control and dependent variables.
For all statistical tests, we used an alpha level of .05, following the common practice in our field to balance the risk of type I and type II errors.

Furthermore, we conducted a qualitative analysis of the interview transcripts, performing the following qualitative coding steps:
(1) We first deductively created an initial codebook covering all our research questions (\autoref{sec:research-questions}).
(2) Two researchers then independently coded the transcripts from the study session of a single group, to find out how well the codebook fits the participants' statements. During this separate coding process, they inductively extended the codebook with new codes to match interesting observations outside the research questions \cite{merriam2015qualitative}. (3) The researchers then discussed and merged their two codebooks into the final codebook:
preference, mental model, scenarios, suggestions/improvements, security, relevance of security in groups, initialization step, verification step, Internet requirement, third party involved, user interface, and user roles.
(4) The two researchers then split the remaining transcripts and deductively coded them according to the final codebook.
(5) Finally, the researchers swapped the transcripts to review and extend the other researcher's codes, and to discuss and resolve coding differences.

\section{Quantitative Results}\label{sec:quantitative}

This section presents our quantitative results and the analysis of participant task performance (\autoref{tab:descriptive-dependent}). The questionnaire includes quantitative scales, numeric fields, and control variables (\autoref{sec:methods-questionnaire}). 
We studied the participants' perceptions of secure contact exchange methods regarding their usability (\autoref{sec:quantitative-usability}), security (\autoref{sec:quantitative-security}), and preference (\autoref{sec:quantitative-preference}).
We further examined their task completion times (\autoref{sec:quantitative-completion}) and general usage patterns for social and collaborative tools (\autoref{sec:quantitative-tools}).
An analysis of the control variables is also provided (\autoref{sec:control-variables}).

\begin{figure}
\begin{subfigure}[t]{0.5\columnwidth}
    %\centering
		\includegraphics[scale=0.78]{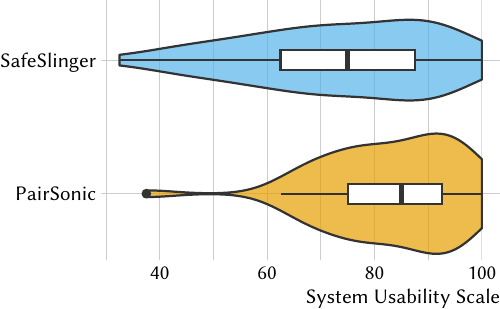}
    \caption
    {
    \gls{SUS} scores.
    }
    \Description{%
    Two violin plots showing the density estimation of the distribution of our participants' SUS scores for \SafeSlinger and \OurApproach, respectively.
    Median values and the interquartile ranges are provided in \autoref{tab:descriptive-dependent}.
    \OurApproach has a higher median SUS score.
    For \OurApproach, there is one outlier at a SUS of 37.5.
	}
    \label{fig:evaluation-sus}
\end{subfigure}%
\begin{subfigure}[t]{0.5\columnwidth}
    %\centering
		\includegraphics[scale=0.78]{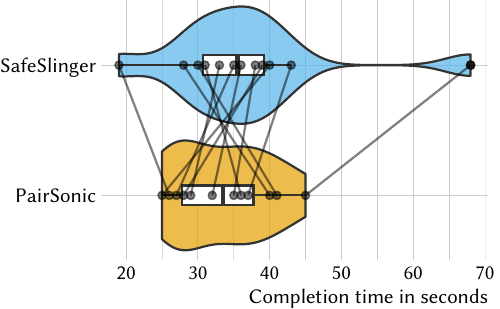}
    \caption
    {
    Completion times.
    }
    \Description{%
    Two violin plots showing the density estimation of the distribution of our participants' completion times for \SafeSlinger and \OurApproach, respectively.
    The violin plot for PairSonic shows a dense clustering around lower time values, suggesting quicker task completion compared to SafeSlinger.
    Median values and the interquartile ranges are provided in \autoref{tab:descriptive-dependent}.
	}
    \label{fig:evaluation-time}
\end{subfigure}
\caption%
{
Comparison of our participants' \gls{SUS} scores and completion times for \SafeSlinger and \OurApproach.
The violin plots show a density estimation of the distributions. The boxplots show quartiles, median, and outliers.
}
\Description{%
A two-part image with two violin plots each showing the SUS scores and completion times for SafeSlinger and PairSonic.
}
\end{figure}

\subsection{Usability}\label{sec:quantitative-usability}
After getting hands-on experience with both group pairing protocols, the participants completed our questionnaire, including the \gls{SUS} \cite{brooke1996sus}.
We use the Shapiro-Wilk test \cite{shapiro1965analysis} to determine whether this interval variable is normally distributed.
The \gls{SUS} scores for \SafeSlinger ($W=0.96$, $p=.095$) are approximately normal, but the scores for \OurApproach ($W=0.92$, $p=.003$) are significantly non-normal distributed.
\autoref{fig:evaluation-sus} depicts the statistics of the \gls{SUS} scores per protocol. Both distributions have visible negative skew, indicating that most participants selected values on the higher end of the scale. Additionally, there is a single outlier for \OurApproach.
Parametric methods such as the t-test assume normal sampling distributions and can give inaccurate results in the presence of outliers \cite{wilcox2017robust}. We, therefore, use non-parametric statistical methods for data analysis (\autoref{sec:methods-analysis}).

\gls{SUS} scores for \SafeSlinger (Mdn = 75) differed significantly from \OurApproach (Mdn = 85) according to the Wilcoxon signed-rank test, $V = 597.5$, $p=.004$. The effect size is $r=-.43$, which corresponds to a large difference according to Cohen \cite{Cohen1992effectsize}.
Thus, \textbf{\OurApproach showed significantly better usability than \SafeSlinger.}

\subsection{Security}\label{sec:quantitative-security}
\begin{figure}
\begin{subfigure}[t]{0.5\columnwidth}
    %\centering
		\includegraphics[scale=0.78]{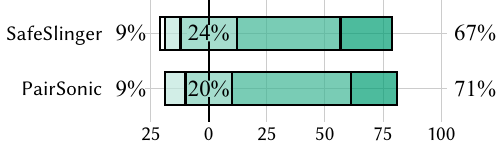}
    \caption
    {
    Security scores.
    }
    \Description{%
        Stacked bar charts showing the security perception scores for SafeSlinger and PairSonic, with both systems having a majority of positive responses, indicating a generally positive security perception.
        Median values and the interquartile ranges are provided in \autoref{tab:descriptive-dependent}.
	}
    \label{fig:evaluation-sec}
\end{subfigure}%
\begin{subfigure}[t]{0.5\columnwidth}
    %\centering
		\includegraphics[scale=0.78]{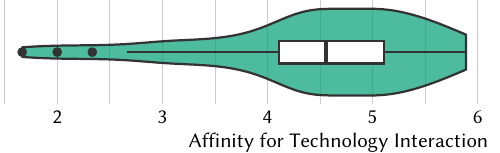}
    \caption
    {
    ATI scores.
    }
    \Description{%

     A violin plot representing the distribution of Affinity for Technology Interaction (ATI) scores among participants, with a median score of 4.6, suggesting a moderately high level of comfort and confidence in using technology.
	}
    \label{fig:evaluation-ati}
\end{subfigure}
\caption%
{
Comparison of our participants' security and ATI scores for \SafeSlinger and \OurApproach.
The stacked bars for the security scores correspond to the five levels of agreement, ranging from  \textquote{strongly disagree} (left) to  \textquote{strongly agree} (right), centered at the neutral response.
The percentages (left, middle, right) represent the share of negative, neutral, and positive responses, respectively.
The violin plot shows a density estimation of the \gls{ATI} distribution. The boxplot shows quartiles, median, and outliers.
}
\Description{%
A two-part image with a vertically stacked bar chart showing the participants' security and a violin plot showing the participants' ATI scores.
}
\end{figure}

Based on their experience during the lab tasks, we asked our participants how secure they think both systems are  on a five-level Likert item.
\autoref{fig:evaluation-sec} shows the security scores, which are ordinal assessments to the statement \textit{\textquote{I think that this system is secure}}, ranging from \textquote{strongly disagree} to \textquote{strongly agree}.
According to the sign test, which is appropriate for ordinal values \cite{siegel1956nonparametric}, the security scores for \SafeSlinger (Mdn = \textquote{agree}) did not differ significantly from \OurApproach (Mdn = \textquote{agree}), $p=1$,
$g=0$.
Hence, \textbf{users have a similar security impression of both methods.}

\subsection{Preference}\label{sec:quantitative-preference}
After our participants tried both systems, we asked them which one they liked better.
\autoref{fig:groupwise-pairsonic-preference} in the appendix shows how many participants preferred \OurApproach in each study group.
Most participants (69\%; 31) preferred \OurApproach compared to the state of the art (SafeSlinger), which is a significant difference according to the binomial test, $p = .016$.
The effect size is $g=.19$, which corresponds to a medium difference according to Cohen \cite{Cohen1992effectsize}.
Overall, \textbf{participants prefer \OurApproach compared to \SafeSlinger.}

\subsection{Completion Time}\label{sec:quantitative-completion}

As an objective measure, we determined the time it took our participants to complete each contact exchange protocol, after a brief learning period during the initial two tasks. \autoref{fig:evaluation-time} shows the completion times of each individual group, connected for both protocols, to show the change within each group.
The completion times for \OurApproach are quite consistent and approximately normal distributed ($W=.94$, $p=.528$). There is an outlier for \SafeSlinger as it took one group \SI{68}{\second} to complete the protocol, making the \SafeSlinger completion times significantly non-normal ($W=.85$, $p=.035$).
This reaffirms our choice for the robust non-parametric Wilcoxon signed-rank test.
Most groups (8; 67\%) were faster with \OurApproach, but the completion times between \SafeSlinger (Mdn = \SI{35.5}{\second}) and \OurApproach (Mdn = \SI{33.5}{\second}) did not differ significantly ($V=49.5$, $p=.432$, $r=-.23$).
The completion times for both schemes showed no significant correlation with the group size ($\tau = .46, p=.052$ in both cases).
In summary, \textbf{the completion times are similar for \OurApproach and \SafeSlinger.}

\subsection{Usage of Social and Collaborative Tools}\label{sec:quantitative-tools}

\begin{figure}
\includegraphics[scale=0.78]{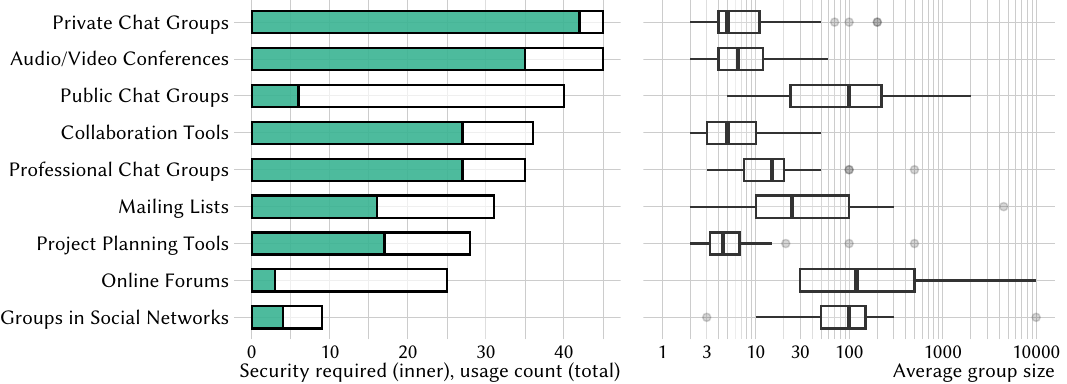}
\caption%
{
The left-hand bar chart illustrates the number of participants using the given social or collaborative tools (total bar length), with the inner green bar representing those who deem security and authentication important.
The right-hand boxplots depict the distribution of the average number of group members indicated by our participants for each tool type. They show quartiles, median, and outliers on a logarithmic axis. One outlier is not included: one participant gave an average online forum group size of 1.6 million.
}
\label{fig:tool-usage}
\Description{%
    A two-part image with vertically stacked bar charts on the left and boxplots on the right showing which services participants use, self-reported security importance for each service, and the estimated average group size of each service. The bar chart shows the count of participants using various types of communication services like private chat groups, audio/video conferences, etc., with an inner green bar indicating the number of users who prioritize security in each type. The boxplots show the distribution of the average group sizes for each communication tool on a logarithmic scale, demonstrating a wide range of group sizes.
}
\end{figure}

We were also interested in understanding which types of social and collaborative tools our participants use, as this would reveal the most prominent use cases.
\autoref{fig:tool-usage} summarizes our participants' responses.
All participants use private chat groups, such as WhatsApp or Signal, as well as audio or video conferencing tools like Zoom. Additionally, a substantial portion partakes in public chat groups (89\%), online collaboration tools like Google Docs (80\%), and professional chat groups such as Microsoft Teams (78\%), showing the relevance of digital collaboration applications.

We also queried our participants about their security desires for these tools, i.\,e., ensuring no unauthorized access.
Tools requiring such security measures could benefit from group pairing protocols, such as \OurApproach, indicating potential use cases.

\autoref{fig:tool-usage} demonstrates that participants seek authentication for most tools, especially private chat groups, audio/video conferencing, online collaboration tools, and professional chat groups.

Futhermore, we asked our participants to estimate the average number of group members for each type of collaboration.
Groups demanding security requirements were typically smaller: the group sizes for private chat groups (Mdn = 5), online collaboration tools (Mdn = 5), and audio or video conferences (Mdn = 6) were much smaller compared to public chat groups (Mdn = 100) and online forums (Mdn = 135).
Our participants are active in a significant number of groups, with a median of 70 groups (IQR = 70).

\subsection{Control Variables}\label{sec:control-variables}

As a control variable, we measured our participant's ATI (Mdn = 4.6, IQR = 1) and smartphone familiarity.
All 45 participants except one have been using a smartphone for more than two years, indicating general familiarity with smartphone usage.

We calculate bivariate Kendall's rank correlation coefficients in \autoref{tab:correlations} to determine the relationship between the control and dependent variables. As shown, our participants' \gls{SUS} scores, security ratings, and preferences do not significantly depend on their technology affinity or smartphone familiarity.
The \gls{SUS} scores for \SafeSlinger significantly correlated with our participant's preference, indicating that participants who found \SafeSlinger less usable preferred \OurApproach instead.
There was also a significant negative correlation between \SafeSlinger \gls{SUS} scores and completion times.
Additionally, we found that a method's \gls{SUS} score significantly correlated with that method's security assessment.

We noticed that the order in which our participants encountered the systems significantly correlates with the \gls{SUS} scores for \SafeSlinger (\autoref{fig:evaluation-sus-depending-order}), but not with the \gls{SUS} scores for \OurApproach or with the other variables.
We discuss this potential order effect in \autoref{sec:limitation-order}.

\section{Qualitative Results}\label{sec:qualitative}

Our previous section's quantitative evaluation indicated that participants found \OurApproach to be more usable and preferred it over \SafeSlinger.
In this section, we complement the quantitative analysis with a qualitative analysis: we conducted post-test interviews with the participants to understand the reasons behind their preference.
We begin by determining the factors contributing to their overall preference (\autoref{sec:qualitative-overall-preference}). Subsequently, we closely compare user perceptions regarding the two main phases of group pairing protocols: the initialization phase (\autoref{sec:qualitative-initialization-preference}) and the verification phase (\autoref{sec:qualitative-verification-step}).
Lastly, we capture our participants' perspectives on acoustic communication (\autoref{sec:qualitative-acoustic}).

\subsection{Which Method Do Users Prefer Overall?}\label{sec:qualitative-overall-preference}

During the interviews, our participants' comments aligned with their preference for \OurApproach overall.
They specifically highlighted two key advantages: better \textit{usability} and \textit{scalability}:

\studyquote{If you have used [\OurApproach{]} once or twice, then it's also just much, much more convenient than [\SafeSlinger{]}. I mean, you open the app, say \guillemotright here, you're the person who sets this up\guillemotleft, hold the phones together, done.}{7}

However, some participants argued in favor of \SafeSlinger, especially for \textit{smaller groups} where the manual comparison tasks require less effort.

Regarding \textit{security}, the qualitative results were mixed, supporting the results of our quantitative analysis (\autoref{sec:quantitative-security}).
Some participants expressed that \SafeSlinger's additional user interaction and complexity instilled a sense of enhanced security compared to the simpler \OurApproach.
Conversely, other participants favored the security of \OurApproach, as it did not solely rely on potentially error-prone user interaction but also incorporated acoustic verification as an additional layer of security.

Most participants raised concerns about the \textit{Internet requirement} of \SafeSlinger, citing instances where they lacked cellular connectivity or access to public WiFi. They worried that this could limit the situations in which they can exchange their contact data.
Additionally, a few participants expressed privacy concerns since the contact data was transmitted over the Internet. Despite \SafeSlinger being designed to safeguard the confidentiality of exchanged information from third parties \cite{farb2013SafeSlinger}, the fact that the protocol relied on Internet connectivity was enough to make some participants worry about their data.
In contrast, the decentralized nature of \OurApproach emerged as a distinct privacy and availability advantage, as no third parties are required to facilitate the protocol.

\subsection{Which Initialization Step Do Users Prefer?}\label{sec:qualitative-initialization-preference}
The initialization step of \SafeSlinger and \OurApproach temporarily associates the group members' devices to exchange contact information and cryptographic public keys.
However, there are distinct approaches employed by each protocol. In \SafeSlinger, a peer-based approach is used, requiring each group member to input the group size and determine the lowest identification number among all members. 
On the other hand, \OurApproach utilizes a coordinator-based approach, where only one group member needs to input the group size, and the rest of the initialization is automated through acoustic communication.
It is important to note that the role of the group leader in \OurApproach does not confer any special privileges and is only relevant to the contact exchange without implications for subsequent communication.

During the interviews, we asked participants to compare the two initialization steps. The majority expressed a preference for our coordinator-based initialization, although some participants argued in favor of the peer-based initialization. This observation is intriguing, particularly when considering a previous user study by Nithyanand et al. \cite{nithyanand2010Groupthink}, which suggested a greater overall acceptance of peer-based methods. This discrepancy invites a deeper analysis to understand the reasoning behind our participants' preferences:

\subsubsection{Effort}
Some participants did not like the additional effort in \SafeSlinger to determine the smallest identification number:

    \studyquote{It is annoying that you have to find out the smallest numbers first. Sure, you can do that relatively efficiently, but if you are in a group with a bunch of people, then maybe people just call in their numbers and then nobody has an overview anymore. }{3}

\subsubsection{Scalability}
Other participants highlighted a scalability issue regarding the effort required, particularly as group size increases:

    \studyquote{If it is a larger group, it is more practical to have a coordinator.}{12}

Participants noted that in \SafeSlinger, all group members are required to enter the group size, which can potentially lead to failures if one member miscounts.

\subsubsection{Determining the Coordinator}
Although no group in our study encountered difficulties in appointing a coordinator during the protocol, some participants expressed concerns about this aspect during the interviews. 
Even though \SafeSlinger's peer-based approach requires more user interaction for each participant, it does not require the explicit social decision of determining the coordinator \cite{jokela2013comparative}.
Interestingly, some participants preferred this trade-off:
    
    \studyquote{I think [SafeSlinger's initialization step] would probably be better, because it leads to fewer disputes about who is now the coordinator. Although, of course, that doesn't matter at all in the end. But in different social groups this can lead to disputes.}{11}
   
Other participants argued that there often already is someone who proposes to exchange contact information in the first place, so there would likely be few disputes about the coordinator in practice:
        
    \studyquote{But if I think about it now, typically someone comes up with the idea of exchanging the contact data.
    That means that in the group there will already be some spokesperson who has crystallized, even if only for this specific situation.
    That means that, okay, I'll say here, I'm the coordinator, who else wants to participate and wants to exchange contacts with me, let's do that. And then I know that if four or five people approach me, okay, we are six people, I can set it up and off we go.
    That is, it actually fits relatively organically into the group dynamic, as I would expect in such a situation.}{4}

\subsection{Which Verification Step Do Users Prefer?}\label{sec:qualitative-verification-step}

In \SafeSlinger, all group members are required to compare and select a matching three-word phrase, whereas \OurApproach mostly automates the verification, with group members only needing to verify that all devices display a green checkmark symbol.
We now compare participants' perceptions of the verification step in the group pairing protocols.

\subsubsection{Acoustic Verification Is Effortless}
Most participants preferred \OurApproach, arguing that it is easier to use:

    \studyquote{For the verification at the end, I think I prefer [\OurApproach{]}, because all you have to do is just make sure there is the green check. So even at that, there's no need to communicate and just raise your phone and show the green check to everyone around. And it's faster that way.}{10}

\subsubsection{Word Phrases Are Prone to Errors}
Some users felt nervous comparing the word phrases and worried about making mistakes.
Others liked the thorough verification in \SafeSlinger, as it made the process feel more secure:

\studyquote{So I think, because I have to look again carefully, that [\SafeSlinger{]} is really secure.}{9}

\subsubsection{Rushing Users}\label{sec:qualitative-rushing-users}
However, many participants mentioned that they did not really check whether all participants have the same word phrases, and only roughly compared the words:

\studyquote{We had the situation where only one person didn't match, but the others did. It's tempting to say right after the first match, okay, yes, it matches, I'll tick the box, great, done. And the other person is then either passed over or I don't know what happens.}{4}

Some rushed the comparison due to the pressure of not wanting to keep the group waiting, particularly in the group scenario:

\studyquote{The reason why you don't take the time to compare accurately is that you just hurry. In a larger constellation, you don't want to be the person because of whom the procedure takes longer, that you need more time to go through the individual words.}{11}

Some participants also rushed with \OurApproach, mistakenly assuming that the green checkmark symbol indicated a successful completion of the protocol, and thus confirmed the dialog without reading the instructions.
In the subsequent discussion, the group proposed using a different symbol or adding a confirmation dialog to mitigate this issue.

\subsubsection{Accessibility}
Some participants mentioned accessibility problems with the word phrases due to non-native speakers, foreign accents, or use of uncommon words. 

\studyquote{The word phrases are basically nonsense in English. That is, it depends on how well it is pronounced, how well it is understood, which again becomes difficult in larger groups. And it also becomes more and more difficult depending on the countries of origin. Red and green as signal colors are actually quite international and easy to distinguish.}{4}
    
\subsection{What Do Users Think of Acoustic Communication?}\label{sec:qualitative-acoustic}
As most users lack prior experience with acoustic communication, our study aimed to gauge their reactions to this new technology.
Our interviews revealed a generally positive sentiment; many participants found its novelty appealing and enjoyed their devices producing melodic sounds.
Past research suggests that introducing fun elements, such as gamification, can encourage users to adopt secure behaviors \cite{herzberg2016Can}.

\studyquote{That you can actually hear the sound is very funny. And it also somehow gives a feeling of, yes, I can hear what's happening. It feels nice to use.}{6}

There are mainly two concerns that our participants had with acoustic communication, specifically related to its security and robustness.

\subsubsection{Security}
Some participants felt the audible acoustic verification enhanced safety, while others doubted its security, fearing potential eavesdroppers.
It is important to note, however, that we designed \OurApproach to maintain security even if acoustic messages are intercepted.

Some technically minded participants questioned the system's security, noting that the acoustic messages always sounded identical:

\studyquote{I think intuitively I would find [\OurApproach{}] a little bit more insecure because you see how they communicate. Or you hear it in this case. If you don't know how it works, what they're doing exactly, I might think, yes, if the smartphones are talking there, then maybe someone can listen in. Or record it.}{12}

Despite the acoustic messages sounding similar, they always encode varying data based on the group participants. Better in-app explanations can address both of these security concerns. Alternatively, a different modulation scheme can be used to produce audibly distinct melodies.

\subsubsection{Robustness and Noise Interference}
During the study, unexpected device interactions led to occasional protocol restarts due to failed acoustic transmissions -- something we hadn't encountered in our pre-study testing.
These failures typically occurred when users placed their devices on the table during acoustic transmission, causing impulsive noise that complicated message reception.

Participants also voiced concerns about the system's performance in loud environments, such as parties or festivals.
Our proof-of-concept implementation is not yet optimized to handle these challenging acoustic environments.
Some participants criticized the inconsistent reliability of audible acoustic communication, expressing a preference for the more reliable \SafeSlinger system.
During our study, \SafeSlinger's performance remained mostly consistent, but this requires a stable Internet connection, which is not available in many scenarios (\autoref{sec:design-comparison-deployability}).
We derive recommendations for future work to improve PairSonic's usability by addressing this criticism (\autoref{sec:future-work}).

\subsection{Alignment with Questionnaire Responses}
We cross-validated whether the interview statements of our participants align with their questionnaire responses in our mixed-methods study and found general agreement between the two data sources.
In the questionnaire, two groups favored \SafeSlinger, two were equally split, whereas the majority (8 of 12) preferred \OurApproach (\autoref{fig:groupwise-pairsonic-preference}), aligning with their interview statements.
For instance, group 11 unanimously preferred \OurApproach in the questionnaire, also rating it higher in \gls{SUS} scores.
During the interviews, every member of group 11 consistently expressed a preference for \OurApproach, arguing for its better usability and scalability.

\section{Discussion}\label{sec:discussion}
We now discuss the main findings of our study.
First, we found users favoring \OurApproach for its simplicity (\autoref{sec:discussion-effortless}), particularly valuing the reduced effort by automating \SafeSlinger's manual verification (\autoref{sec:discussion-automation}).
Interestingly, though, some users perceived the more complex system as more secure (\autoref{sec:discussion-complex-secure}).
We address the associated issue that complex systems tend to be more error-prone (\autoref{sec:discussion-errors}) and investigate how increased transparency and education may enhance perceived security, even for simpler systems (\autoref{sec:discussion-transparency}).
We discuss the applicability of PairSonic for different usage scenarios within the CSCW community based on our participants' comments (\autoref{sec:discussion-scenarios}), alongside a brief security analysis (\autoref{sec:discussion-security}).
We then discuss the scalability of group pairing protocols and compare which actions users have to perform in PairSonic and SafeSlinger (\autoref{sec:discussion-scalability}).
This section concludes by addressing PairSonic's requirement that the users have to physically meet (\autoref{sec:discussion-meetings}), and proposing directions for future research (\autoref{sec:future-work}).

\subsection{Users Generally Prefer Simple and Effortless Group Pairing Systems}\label{sec:discussion-effortless}
We designed \OurApproach to require less user effort than \SafeSlinger, in line with our research hypothesis that users would prefer a simpler, more effortless system (\autoref{sec:research-hypothesis}).
Our lab study supports this hypothesis, as participants significantly favored \OurApproach over \SafeSlinger, giving it higher \gls{SUS} scores (\autoref{sec:quantitative}).
When we interviewed our participants, most explained that they prefer \OurApproach because it is easier to use and requires less effort.
They especially liked the automated verification step, which only requires a simple binary comparison. Furthermore, the group-wise user interaction of moving the devices closer for pairing worked very well, confirming previous research on intuitive device association gestures \cite{chong2013How,jokela2015Connecting,jokela2014FlexiGroups,kray2010Userdefined}.

\subsection{Automating Verification Tasks Using Acoustic Communication}\label{sec:discussion-automation}
Previous user studies have shown that users find it difficult to perform authentication ceremonies with current \gls{E2EE} tools like Signal or WhatsApp, because they are time-consuming and require much effort \cite{herzberg2016Can,vaziripour2017that}. 
Our system, \OurApproach, improves the state of the art because it automates manual verification tasks such as comparing phrases, digits, or pictures.
Users responded favorably to this streamlined approach in our interviews, with most participants being able to exchange contact information in less than \SI{35}{\second}.

Our system utilizes an acoustic \gls{OOB} channel for automated verification, which offers several unique advantages over other wireless channels that modern smartphones support: unlike WiFi or Bluetooth, it is limited to close proximity. This enhances usability by reinforcing the intuitive gesture of bringing devices closer together to share contacts, and also improves security by making distant attacks more challenging.
In comparison to \gls{NFC} or \gls{QR} codes, the acoustic \gls{OOB} channel can easily accommodate multiple devices -- a necessary feature for group scenarios.

Moreover, the acoustic \gls{OOB} channel is a software-defined physical layer, unlike WiFi or Bluetooth whose physical layers cannot be customized in smartphones. This adaptability enables advanced physical layer security techniques, which can protect the group pairing protocol against sophisticated attacks, adding an extra layer of security \cite{putz2020AcousticIntegrityCodes}.
This reduces the need for meticulous user behavior during verification.
By safeguarding the acoustic channel at the physical layer, we can further enhance both security and usability of our system, while meeting all our deployability requirements (\autoref{sec:problem}).

\subsection{Complex and Effortful Systems Can Feel More Secure}\label{sec:discussion-complex-secure}
While most participants favored \OurApproach, our quantitative analysis revealed no significant difference in how users perceive the security of \OurApproach and \SafeSlinger (\autoref{sec:quantitative}).
In fact, the interviews even indicate that some users have less trust in \OurApproach due to its simplicity.
They felt that the added complexity of \SafeSlinger made the system seem more secure:

\studyquote{I think the perceived security with [\SafeSlinger{}] was higher because, well it sounds stupid, but because you enter so many things. [...] But I thought, because it took much longer, that it's probably more secure.}{4}

Our research hypothesis emphasizes minimal user interaction, but our interviews revealed a potential issue for security systems like \OurApproach.
Its effortless nature can make it seem too simple to be secure, almost like magic, leading to distrust.
When a system automatically handles security, users may feel a loss of control, viewing the system as a black box that they can't understand \cite{ruoti2013Confused,ruoti2016We,lerner2017Confidante}.
They might also equate complexity with security due to everyday experiences like complex password requirements or cumbersome multi-factor authentication  \cite{kainda2010Two}.

This unexpected finding emphasizes a dilemma:
a system needs to be simple and effortless for usability, but if it's too simple and requires minimal effort, users may doubt its security.
Our results highlight the challenging trade-off that security engineers must balance -- not just optimizing usability, deployability, and security, but also the \textit{perceived} security.

\subsection{Error-Prone Collaborative User Interaction}\label{sec:discussion-errors}

User interaction requiring substantial effort often leads to errors. This poses a particular problem for systems like \SafeSlinger, which depend on meticulous user behavior for security, such as the verification step involving comparison of word phrases. We noticed in our study that many participants did not carefully compare SafeSlinger's word phrases on each device -- often checking only the first word, a behavior referred to as \textquote{rushing users}. This tendency has also been observed in a previous user study by Tan et al. \cite{tan2017Can}.

As another example, in some larger groups, our participants only compared SafeSlinger's words with adjacent group members and accepted the matching word phrase without confirming this with the rest of the group.
In other groups, a single member took charge of coordinating the verification step.
However, once a few members confirmed a match, the group often selected that phrase based on majority agreement, potentially neglecting quieter members who did not share the same phrase.
Differently to \SafeSlinger, the security of \OurApproach does not heavily depend on user diligence, thus offering systematic security advantages against active adversaries.

In contrast to the traditional pairwise authentication schemes described in \autoref{sec:related-work}, in our collaborative situation the whole group works together to achieve a common security goal.
We made the interesting observation that some users checked and helped each other, leading to fewer overall failures.
This phenomenon was also observed in previous user studies in collaborative scenarios \cite{kainda2010Two,jokela2015Connecting}.

However, Kainda et al.\ found that the primary source of failure was poor communication among group members \cite{kainda2010Two}.
We noted similar issues in some groups, particularly where participants were more reserved and hesitant to voice problems or discrepancies in word phrases.
These findings highlight that a system requiring less user interaction and communication results in a more reliable and less error-prone solution.

\subsection{Transparency and Comprehensibility Can Improve Perceived Security}\label{sec:discussion-transparency}
Technical users typically like to understand how and why a system works, whereas non-technical users need convincing evidence that a system is secure, without necessarily needing to understand all technical details \cite{fassl2021Exploring}.
Our interviews indicated that most users were not familiar with acoustic communication, especially if it is audible, and would benefit from better explanations within the \gls{app}.
Systems like \SafeSlinger, which depend on third parties over the Internet, could also enhance user experience with increased transparency and education regarding data access, addressing users' privacy concerns.
CSCW research during the COVID-19 pandemic has shown that privacy-related trust in the application provider plays an important role in the user's willingness to use online collaborative tools \cite{namara2021differential}.

\subsection{Usage Scenarios}\label{sec:discussion-scenarios}
There are a number of collaboration scenarios that have been proposed by CSCW researchers, where we argue that PairSonic could improve the usability of the group formation process.
Previous research focused mainly on remote collaboration using tools like groupwork platforms \cite{namara2021differential}, crowdsourcing \cite{harandi2019supporting}, and mobile devices \cite{zheng2017colladroida}. 
Additionally, there are separate lines of work studying computer-supported collaboration for creative group work and education~\cite{chen2019group}, for networking and meeting new collaborators \cite{wei2016grouplink,robb2016wellconnected,umbelino2019prototeams}, and within the context of spontaneous ad-hoc interaction \cite{edwards2002using,defreitas2015group}.
A common challenge in these studies is connecting people and their devices, often using technologies like QR codes that struggle with usability and scalability in larger groups (\autoref{sec:current-pairing-insufficient}).
PairSonic offers a scalable, user-friendly drop-in replacement to these legacy pairing technologies, providing a more efficient solution for associating devices in group settings.

There are also usage scenarios potentially involving larger groups of participants.
For example, Tolmie et al.~\cite{tolmie2014supporting} and Fosh et al.~\cite{fosh2016supporting} explored collaborative interactions in cultural sites such as museums, suggesting that connecting visitors, possibly in ad-hoc groups with a tour guide, enhances engagement with both the group and the exhibits.
PairSonic can facilitate this by connecting visitors through their own devices, avoiding the need for museum rental devices.
Our testing confirms PairSonic's effectiveness for larger groups, including those with more than 10 participants, when their devices are close and in low-noise environments.
Museums are ideal for PairSonic, especially using the inaudible frequency range instead of audible acoustic signals (\autoref{sec:implementation}).
The quiet setting ensures reliable acoustic communication, and its inaudibility ensures it does not disturb other visitors.
Additionally, like QR codes, the acoustic channel can transmit small data packets, which could further improve engagement with exhibitions. For example, exhibitions could emit an audio beacon offering additional information to nearby visitors, benefiting from the deployability advantages of the acoustic channel compared to alternatives such as Bluetooth or WiFi (\autoref{sec:design-comparison-deployability}).

Our study participants identified several scenarios where PairSonic could serve as a secure method for exchanging contacts, such as private Signal chat groups, Zoom video conferences, and online collaboration tools (\autoref{sec:quantitative-tools}).
These situations usually have higher security requirements, often relating to personal, familial, or professional collaborations.
Many interviewees shared their preferences for online security in social contexts:

\studyquote{For me, it depends on how many people are in the group and who is in the group. If it's a group where I'm only with friends, where maybe more sensitive data is discussed, then [security] is the most important requirement of all, I would say.}{12}

Others felt knowing the other participants is essential for effective online collaboration:

\studyquote{I find it very awkward to write in groups where I don't know who else can read it. That's why I usually just don't participate at all.}{6}

Our results align with previous CSCW research on the security and privacy needs of users, which found that especially vulnerable types of users and users discussing sensitive topics have a substantial need for interpersonal trust and knowing the other participants \cite{shusas2023accounting}.
Similarly, our participants also named specific scenarios where they saw the potential for securely exchanging contact information using PairSonic. These included festivals, exhibitions, conventions, conferences, but also everyday situations when meeting new people, like meeting new people through friends.
Some expressed interest in using \OurApproach with non-technical users, anticipating potential difficulties these users might face with \SafeSlinger:

\studyquote{You can also use [\OurApproach{}] with people who normally would need further assistance. Instead it's enough if one person is the coordinator, and then the others just have to click yes or no. I think that's actually quite good, because it takes the hurdle out for people who are not so tech-savvy, if you tell them that they don't have to do anything except click yes or no once.}{6}

\subsection{PairSonic Inherits SafeSlinger's Strong Security}\label{sec:discussion-security}
While our primary focus is on the usability of \OurApproach, and a comprehensive security analysis is not within this work's scope, we do provide a brief discussion on its security aspects. We adhered to the security best practice of not unnecessarily creating an entirely new cryptographic protocol from scratch, opting instead to rely on the established security guarantees of the \SafeSlinger protocol whenever possible.
Consequently, \OurApproach protects against all types of attacks mentioned in the \SafeSlinger paper, such as threats from malicious bystanders, impersonation, sybil/hidden node attacks \cite{douceur2002Sybil,kainda2010Two}, and Group-in-the-Middle attacks \cite{kuo2008Mind}.

Unlike the \SafeSlinger protocol, \OurApproach automates information exchange stages that previously required manual effort, using the acoustic out-of-band channel for this purpose.
The acoustic channel must have the same security properties as the human-mediated channel in order to not degrade \OurApproach's security.
Our adversary model assumes the acoustic channel's authenticity, implying that adversaries cannot tamper with it. This assumption is supported by previous studies on wireless authentication, which highlight the physical constraints of sound waves, such as their significantly shorter range compared to radio communication and their inability to penetrate physical barriers like walls \cite{soriente2008HAPADEP,balfanz2002Talking,stajano2000Resurrecting}.
Furthermore, the authenticity of the audio channel can be explicitly ensured using physical layer security techniques  \cite{putz2020AcousticIntegrityCodes}, whose assumptions and system model align with \OurApproach's protocol. Future work could incorporate them to enhance or replace the current ggwave physical layer used in our protocol.

In our adversary model, we assume that an adversary can eavesdrop on the audio channel, enabling them to gain information about the ad-hoc WiFi network and potentially interfere with it.
However, any WiFi interference attempt by the adversary to alter the group's contact information would be futile, as \OurApproach is designed to detect such modifications and terminate the protocol during the acoustic verification phase.
The use of the acoustic out-of-band channel in \OurApproach also enhances security by supporting longer hash values than \SafeSlinger, reducing the likelihood of hash collisions.
The \SafeSlinger protocol employs 24-bit hash values as Short Authentication Strings. Increasing the entropy of these hashes would negatively impact usability, as it would require users to manually compare a greater number of words or more complex phrases. In contrast, \OurApproach streamlines this process by automatically transmitting the hash value via the acoustic channel. This method allows for longer hash values without user involvement and is less susceptible to security-critical user errors, a concern highlighted in our study (\autoref{sec:discussion-errors}).

\subsection{Scalability}\label{sec:discussion-scalability}

\begin{figure}
\includegraphics[scale=1.05,page=1]{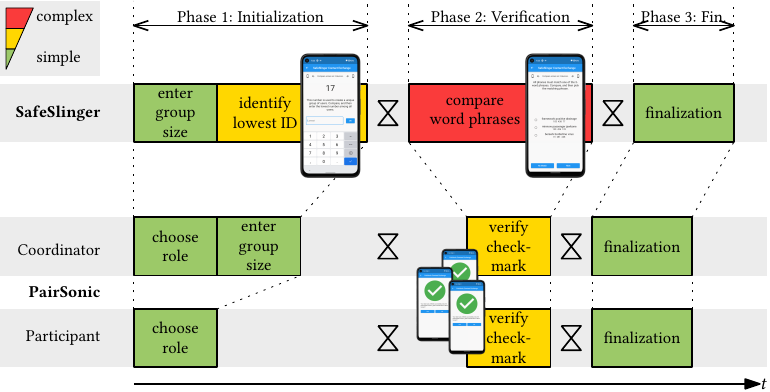}
\caption%
{
Qualitative timelines show which actions users have to perform during all three phases of the contact exchange process (\autoref{tab:protocol-comparison}). The symbol \includegraphics[scale=0.5,page=2]{gfx/scalability-timeline-export.pdf} indicates waiting time. The yellow and particularly the red user actions are more complex than the green actions and take more time for larger groups, showing that \OurApproach scales better than \SafeSlinger.
}
\label{fig:scalability-timeline}
\Description{%
    A comparative timeline showcasing the actions users must perform during the contact exchange process with SafeSlinger and PairSonic. Each step is color-coded for complexity, with SafeSlinger having more complex (yellow and red) steps such as identifying the lowest ID and comparing word phrases, compared to PairSonic's simpler (green) steps. PairSonic simplifies the process, with the coordinator choosing a role and entering the group size while participants only choose a role. Verification is simplified in PairSonic with automatic acoustic verification, leading to a quicker and simpler finalization step for all roles.
}
\end{figure}

Besides usability, one motivation for \OurApproach was improved scalability. 
\OurApproach requires less interaction between participants by design, making it better suited for larger groups of users than \SafeSlinger.
To illustrate this, \autoref{fig:scalability-timeline} shows timelines of \OurApproach compared to \SafeSlinger for a typical user group during our study.
From a user's perspective, each protocol consists of phases where they have to wait, but also phases requiring them to read, choose an option, compare information, or enter a number: these are phases requiring manual actions. The total cumulative duration that users have to diligently perform these actions (i.e., excluding waiting times) is lowest for \OurApproach with the participant role, followed by the coordinator role, with \SafeSlinger requiring the longest attentive time for all users.

The figure also highlights the protocol phases that take noticeably more time for larger groups.
\SafeSlinger contains two such phases: identifying and entering the lowest identification number amongst all participants, and the word phrase comparison. \OurApproach, however, only has a single phase scaling with the group size, namely determining that every device shows a large checkmark symbol, a task that is comparatively much simpler than the tasks in \SafeSlinger, since it only requires a quick glance at the smartphones.
In summary, \OurApproach requires less attentive user duration, making it well-suited also for larger groups.

Larger groups further reinforce the issue of error-prone collaborative user interaction that we discussed previously (\autoref{sec:discussion-errors}).
With \SafeSlinger, larger group sizes involve more manual text comparisons, which not only raises the likelihood of mistakes, but also reinforces the problem of \emph{rushing users}. We observed that 
some users felt pressured by larger groups and wanted to avoid keeping the group waiting (\autoref{sec:qualitative-rushing-users}). \OurApproach was less affected by group size, but our proof-of-concept implementation struggled with acoustic noise interference (\autoref{sec:qualitative-acoustic}), which might occur more often with more participants. 

Contrary to our expectation, we did not find a significant difference in completion times between both protocols, even though most groups in our study were faster with \OurApproach.
The completion times of \SafeSlinger were faster than expected and can be attributed to many participants not thoroughly comparing the word phrases on each device, as explained earlier (\autoref{sec:discussion-errors}).
While this explains the comparable completion times, it degrades \SafeSlinger's security in practice, as partial comparisons involve less entropy \cite{tan2017Can}.
We also note that the scalability advantage of \OurApproach is likely more pronounced for participants who take longer with \SafeSlinger's word comparison, such as older users, users with bad eyesight, or shy participants.

\subsection{Physical Meetings}\label{sec:discussion-meetings}
PairSonic assumes that all users are co-located, making it suitable for offline and online collaboration, provided the users meet in-person first (\autoref{sec:discussion-scenarios}).
This limitation is by design and arises from our requirement that the system functions without relying on any central infrastructure and without necessitating any pre-established security context (\autoref{sec:problem-definition}).
While SafeSlinger can operate remotely over the Internet, it violates these requirements by assuming the existence of another trusted communication channel to authenticate each other.
However, such a channel typically does not exist yet, which is precisely why the group pairing process is needed to begin with.

This requirement for physical meetings means that to securely collaborate with a new contact using PairSonic, you must first meet them in person.
In larger collaborative groups, it can often be impractical to meet with every new member as they join the group.
However, PairSonic should be seen as just one component of a broader key exchange process, which cannot address the whole key management lifecycle by itself \cite{blaze1996Decentralized}. 
To address this limitation, PairSonic could be integrated into a more comprehensive key management system like the decentralized web of trust (e.g., OpenPGP \cite{rfc4880}). This integration could replace the traditional manual verification of PGP keys in person or as part of keysigning parties \cite{caronni2000walking,ulrich2011investigating}.

In such a system, transitive trust can be utilized, meaning that you might choose to trust the contact list of someone you have already verified, rather than verifying every new contact yourself. 
As a result, a new group member would only need to meet and exchange keys with one existing group member. This member then signs and forwards the authenticated contact information to the rest of the group. In this scenario, the initial group members use PairSonic to establish an initial security context, which then serves as a base for other key management protocols that do not require physical interaction.\footnote{Similar established strategies could be employed to deal with other aspects of the key management lifecycle, such as the loss or change of keys.
For example, OpenPGP typically manages key replacement by generating a new keypair, signing it with the old key, and then issuing a revocation certificate for the old key.
This process ensures a seamless transition from the old to the new key while maintaining security and trust within the system.}

\subsection{Future Work}\label{sec:future-work}
Our study identified the primary weakness of \OurApproach as the limited robustness of the acoustic \gls{OOB} channel in noisy environments, reducing the potential use cases for contact exchange.
Whereas our paper focused on the usability evaluation of \OurApproach's prototype, the interviews have shown that future work is needed to improve the \gls{OOB} channel's reliability concerning realistic noise sources, exploring both audible and inaudible frequency ranges.
Additional future work should include a comprehensive evaluation of the acoustic channel's functionality and performance for various environments, to determine the practical limits of this novel technology.
The acoustic physical layer in the user study by Mehrabi et al.~\cite{mehrabi2019Evaluating} was more reliable, suggesting that a better design can address this issue.

Additionally, enhancements to the user interface could further increase \OurApproach's usability, reflecting findings from previous studies on other authentication ceremonies \cite{vaziripour2018Action,wu2019Something}.
During our study, we observed that some participants rushed the verification step in \OurApproach.
This was not to bypass a tedious task, as seen in previous approaches, but because they misconstrued the green checkmark symbol as a sign of successful completion (\autoref{sec:qualitative-verification-step}).
Using a different symbol, such as a question mark, could signal the need for user involvement in security verification, potentially resolving this issue.
Alternatively, a confirmation dialog could be introduced following the initial prompt to avoid unintentional acceptance.
Future research could validate these enhancements by complementing our lab study with an additional field study analyzing the long-term usage of group pairing protocols in real-world conditions.

Lastly, tech-savvy users expressed curiosity about \OurApproach's operation, its security properties, and how the acoustic channel protects security. Including an in-app help menu with detailed explanations could assist users with questions about the system's functionality or security.
Similarly, Herzberg et al.~\cite{herzberg2016Can} suggested that for an authentication ceremony to be effective, it must not only be usable, but users also need to understand its necessity.

\section{Limitations}
This section addresses the limitations of our work due to our recruitment process and study design.

\subsection{Recruitment}
Our sample was relatively young, mainly comprising students and participants aged 20-29. Rather than selectively sampling for a representative distribution, we controlled for factors potentially influencing usability (\autoref{sec:control-variables}). Neither our control variables nor age significantly correlated with usability, security, or preference scores.

However, a previous study by Kobsa et al. \cite{kobsa2009Serial} reported significant differences in completion times between younger and older participants in pairing protocols.
Given this, our younger sample may underestimate completion times for the broader population. Particularly, the scalability difference between \SafeSlinger's word comparison and our automated verification could be more notable among older users.

\subsection{Order Effect in Our Within-Subjects Design}\label{sec:limitation-order}
In our correlational analysis (\autoref{sec:control-variables}), we found some evidence of a potential order effect impacting the \gls{SUS} scores. Participants who first tried \OurApproach generally rated \SafeSlinger lower (Mdn = 67.5), possibly due to the extra effort \SafeSlinger requires (contrast effect).
However, we found no evidence of a contrast effect in reverse: participants who started with \SafeSlinger typically assigned high initial \gls{SUS} scores (Mdn = 86.25), likely because they had to fill in the \gls{SUS} questionnaire before trying the second system.
Although these participants generally rated \OurApproach higher after using it, the high initial score for \SafeSlinger limited the relative improvement.
We counterbalanced the sequence of systems in our study design to minimize potential order effects. Regardless of the order of exposure to the systems, participants generally gave \OurApproach higher \gls{SUS} scores.

\subsection{Anchor Effect in Groups}
We conducted our study with groups of multiple participants.
During the group interviews, we sequentially asked the same question to each participant, which may have triggered an anchoring effect, as participants' responses could have influenced each other.
We noticed that some participants picked up arguments from previous respondents, but we also noticed instances of disagreement; most groups did not have a unanimous opinion (\autoref{fig:groupwise-pairsonic-preference}).
Our study design partially mitigates the anchoring effect by having participants complete questionnaire sections before the corresponding interviews.
This approach allowed them to reflect on the tasks and form their own opinions prior to the discussion.
Given the cooperative nature of the group pairing process, conducting the experiment within a group of participants appears justified.

\subsection{Lab Study}
Our participants tested PairSonic and SafeSlinger three times each in our lab. Although a lab setting does not fully replicate real-world conditions \cite{chong2014Survey}, it offers a consistent and reproducible environment, which minimizes confounding factors.
While a field study might have more accurately represented real-world usage, the presence of uncontrollable environmental and external variables could have adversely impacted the quality of the data.
Given that PairSonic is still a proof-of-concept implementation, the lab environment was particularly advantageous for observing our participants' interactions and reactions to the apps in real time and to discover problems in the user experience.
While observing participants might influence their behavior, this setup enabled us to identify an unexpected user interaction that resulted in protocol failures (\autoref{sec:qualitative-acoustic}). Such issues, which were not evident during our testing, would have been challenging to detect in a field study.
In addition, our study design strategically interleaved practical tasks with post-test questionnaires and interviews. This allowed us to promptly capture our participants' thoughts on each system before they experienced the other, a process more feasibly managed in the lab setting. We also wanted to specifically evaluate how our participants would handle potential active attacks on the pairing process. Simulating such attacks in a controlled manner is much easier in a lab environment.

\section{Conclusion}
We introduced a novel, practical method for secure contact information exchange, known as \textit{\OurApproach}, and compared it to the state-of-the-art system \SafeSlinger \cite{farb2013SafeSlinger} in a within-subjects lab study ($N=45$).
Our system's primary advantage is that it reduces user effort by automating laborious manual tasks using acoustic communication.
Our results reveal that while users value ease in authentication systems, they often associate more complex systems with higher security.

\briefSectionBF{RQ1}
\textit{\textquote{Which initialization and verification steps do users prefer?}}
Our participants showed a significant preference for \OurApproach over \SafeSlinger.
In the interviews, most preferred our leader-based initialization method and the acoustic verification step.

\briefSectionBF{RQ2}
\textit{\textquote{Which method has better usability?}}
\OurApproach is significantly more usable than \SafeSlinger.
Our participants appreciated the seamless contact exchange and the effortless scalability to larger groups. However, there's a need to enhance the reliability of the \gls{OOB} channel.

\briefSectionBF{RQ3}
\textit{\textquote{How do users perceive the security of both methods?}}
We observed no significant difference in the security ratings for both methods.
Although minimizing user interaction improved usability, it surprisingly decreased perceived security for some participants, according to our interviews. Many users, drawing from their experiences, associate security with complexity, underscoring a tricky trade-off between usability and perceived security.

\briefSectionBF{RQ4}
\textit{\textquote{How do users like the audible acoustic \gls{OOB} channel for pairing?}}
Our study indicates that acoustic communication is a promising, user-friendly method for data exchange between nearby devices, reinforcing the intuitive action of bringing devices close for association.
However, future work should focus on enhancing the robustness of the physical layer in everyday scenarios.

\section*{Availability}
Together with this paper, we provide an overview of the PairSonic project online at \url{https://seemoo.de/s/pairsonic}.
PairSonic is available as open-source software on GitHub:  \url{https://github.com/seemoo-lab/pairsonic}. The PairSonic app will also be presented as a demo at CSCW 2024 \cite{putz2024pairsonicdemo}.

Alongside this paper, we release a replication package that includes our evaluation scripts and the pseudonymized dataset from our study \cite{zenodo2024dataset}.
This dataset contains usability, security, and preference scores, completion times, reported usage of nine types of social and collaborative tools, and seven demographic and control variables, for each of our 45 participants.

\begin{acks}
This work has been funded by the LOEWE initiative (Hesse, Germany) within the emergenCITY center [LOEWE/1/12/519/03/05.001(0016)/72].
We thank the anonymous reviewers for their helpful suggestions.
Furthermore, we thank Maximilan Gehring for his contributions to the initial PairSonic prototype, and Lea Holaus for drawing the illustration in \autoref{fig:teaser}.
We also acknowledge SafeSlinger's substantial innovation over previous pairwise systems, which greatly inspired us in developing PairSonic.
\end{acks}

\bibliographystyle{ACM-Reference-Format}
\typeout{} % fix for overleaf bib rendering
\bibliography{bibliography}

\appendix

\clearpage
\section{Supplementary Figures}\label{sec:supplementary-figures}

\begin{figure}[!h]
    %\centering
		\includegraphics[scale=0.78]{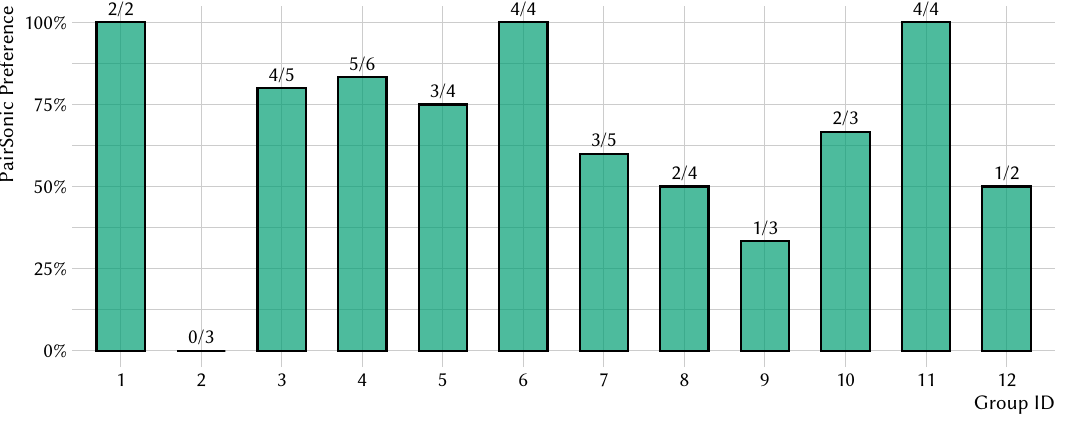}
\caption%
{
This bar chart shows the proportion of group members preferring \OurApproach in each study group. The label on top of each bar shows the corresponding number of group members who preferred \OurApproach and the total group size.
Only four groups had a unanimous preference (three favoring \OurApproach, one favoring \SafeSlinger). This variation suggests the possible anchor effect discussed in \autoref{sec:limitation-order} might not be very strong.
}
\label{fig:groupwise-pairsonic-preference}
\Description{%
    A bar chart indicating the proportion of participants in each study group who preferred PairSonic. Each bar represents a different group, labeled by group ID, with the height reflecting the percentage of participants who favored PairSonic. Numerical ratios above each bar show the count of participants preferring PairSonic out of the total group size. The chart shows variability in preference across groups, with three groups unanimously preferring PairSonic and others showing a mixed preference.
}
\end{figure}

\begin{figure}[!h]
\begin{subfigure}{0.25\columnwidth}
    \centering
		\includegraphics[width=\columnwidth]{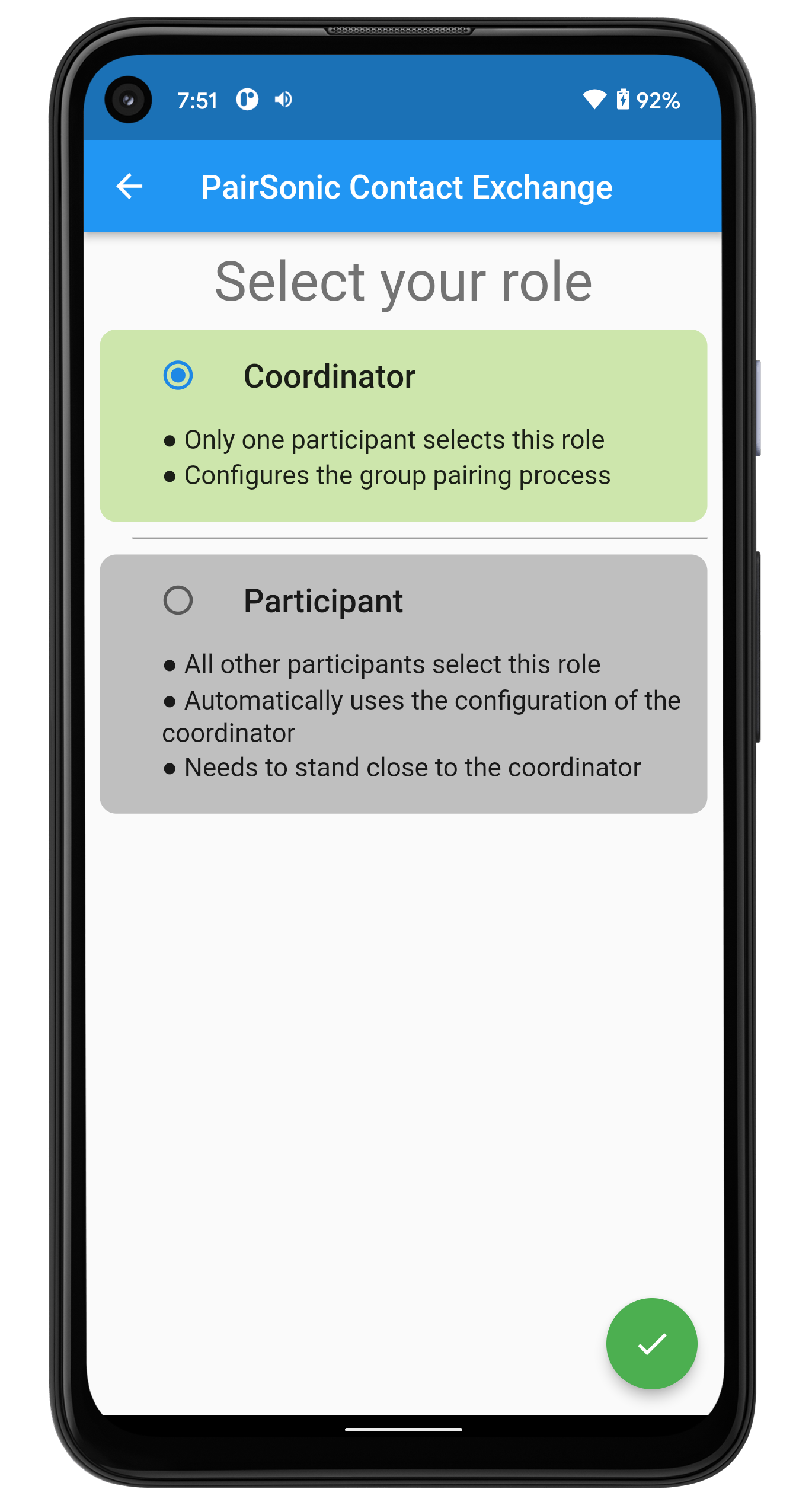}
    \caption
    {
        Role selection.
    }
    \Description{%
        Role selection screen with two buttons where the coordinator is selected.
	}
    \label{fig:pairsonic-coordinator1}
\end{subfigure}%
\begin{subfigure}{0.25\columnwidth}
    \centering
		\includegraphics[width=\columnwidth]{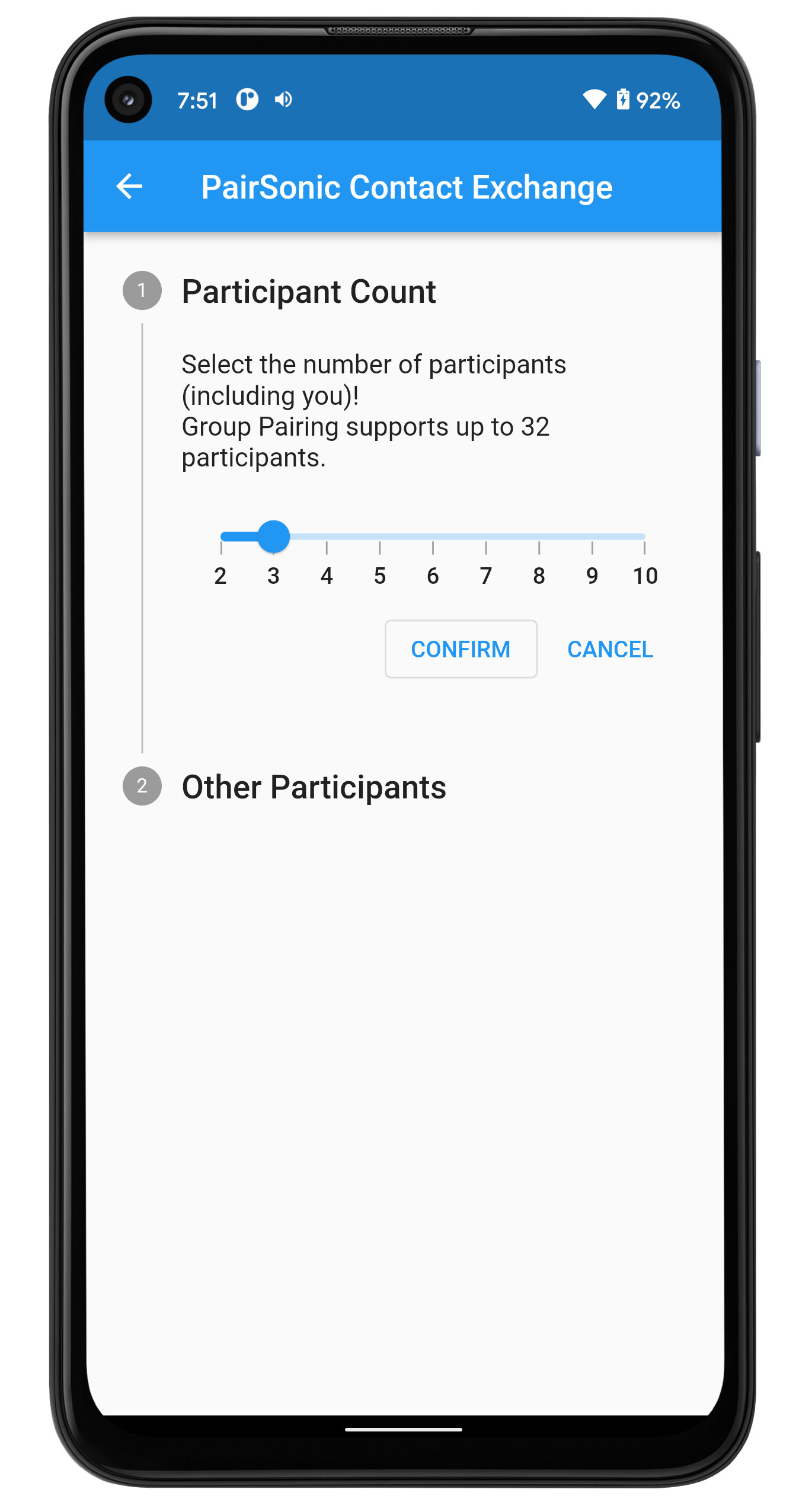}
    \caption
    {
        Participant count.
    }
    \Description{%
        Screen to input the number of participants in the group with a slider interface.
	}
    \label{fig:pairsonic-coordinator2}
\end{subfigure}%
\begin{subfigure}{0.25\columnwidth}
    \centering
		\includegraphics[width=\columnwidth]{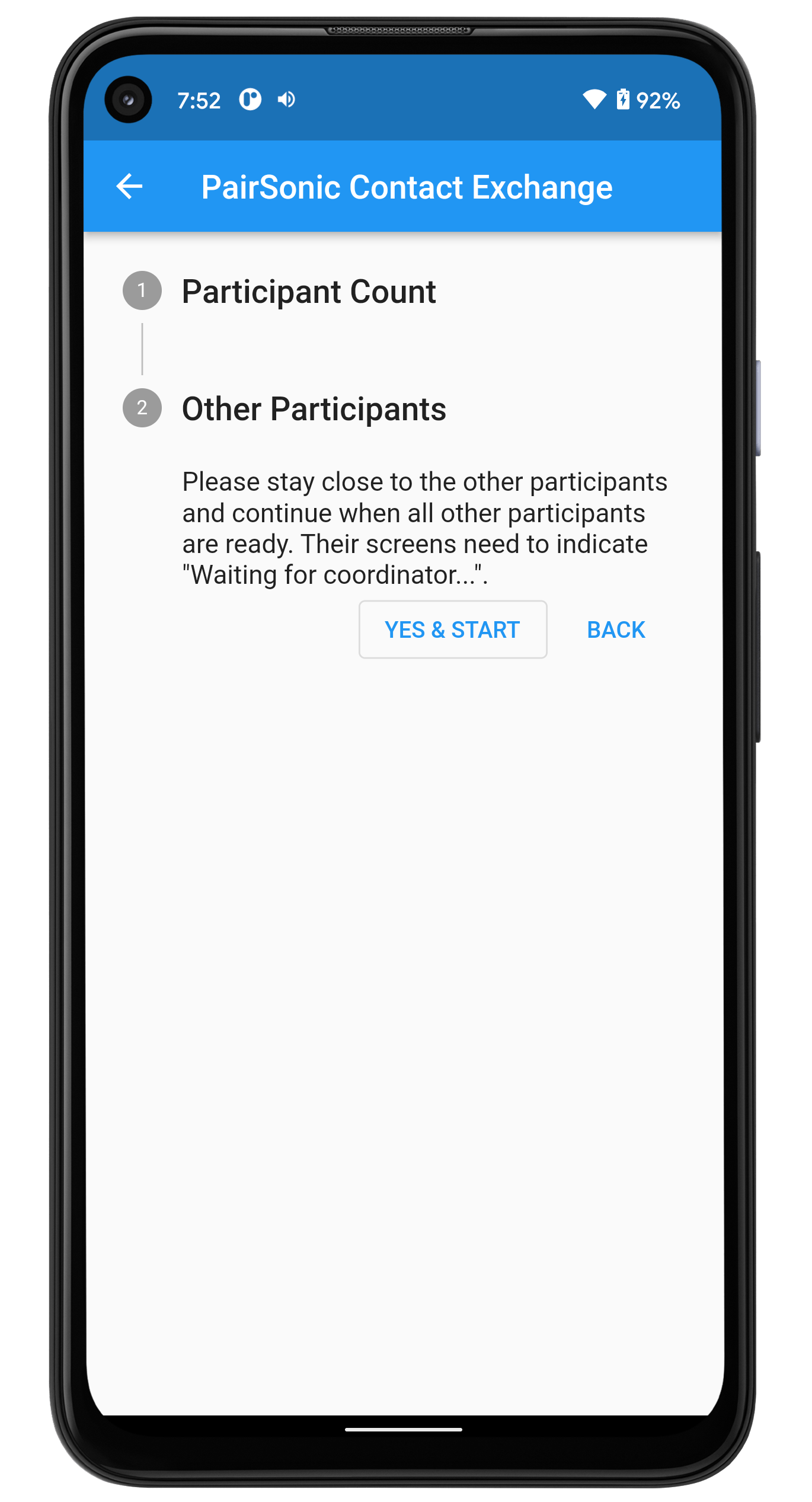}
    \caption
    {
        Ready check.
    }
    \Description{%
        Ready check screen prompting the coordinator to ensure all participants are prepared to start.
	}
    \label{fig:pairsonic-coordinator3}
\end{subfigure}%
\begin{subfigure}{0.25\columnwidth}
    \centering
		\includegraphics[width=\columnwidth]{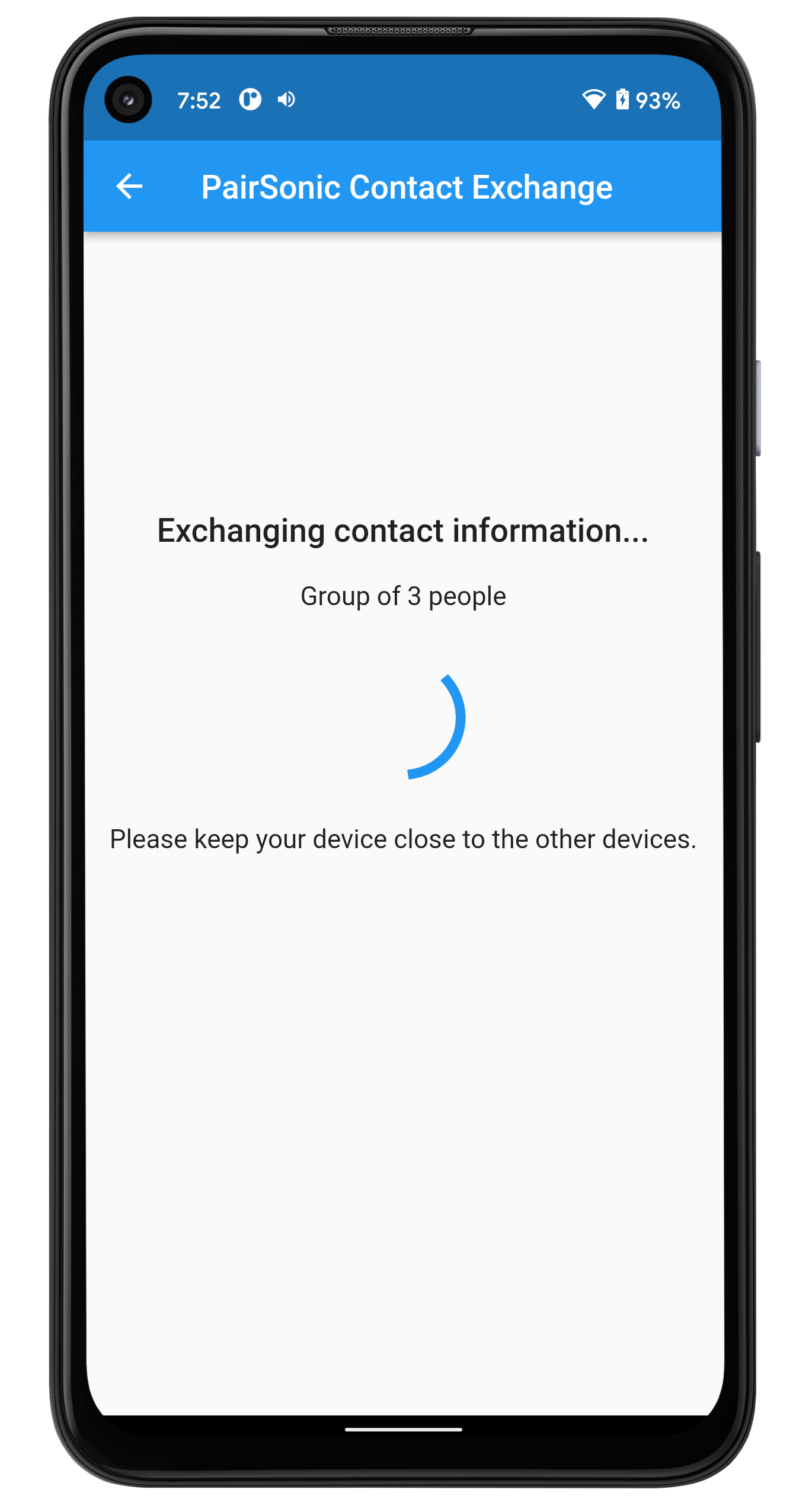}
    \caption
    {
        Exchange begins.
    }
    \Description{%
        The final screen of the initialization process indicates the beginning of the exchange process, with instructions to keep the device close to the others with a progress circle element.
	}
    \label{fig:pairsonic-coordinator4}
\end{subfigure}
\caption%
{
This figure depicts \OurApproach's initialization phase from the perspective of the \textit{coordinator} role.
(a) The coordinator begins by selecting their role.
(b) They proceed to select the total number of participants, counting themselves.
(c) Before continuing, they ensure that the remaining participants are in the ready state (\autoref{fig:pairsonic-participant2}).
(d) The coordinator's smartphone then initiates the exchange via the acoustic \gls{OOB} channel. The verification and finalization phases that follow are identical for both the \textit{coordinator} and \textit{participant} roles, as shown in \autoref{fig:pairsonic-participant3} and \autoref{fig:pairsonic-participant4}.
}
\label{fig:ourapproach-smartphone-flow-coordinator}
\Description{%
    A four-part image with screenshots displaying the PairSonic app's initialization phase, from the coordinator's viewpoint.
}
\end{figure}

\begin{figure}[!h]
\begin{subfigure}{0.5\columnwidth}
    %\centering
		\includegraphics[scale=0.78]{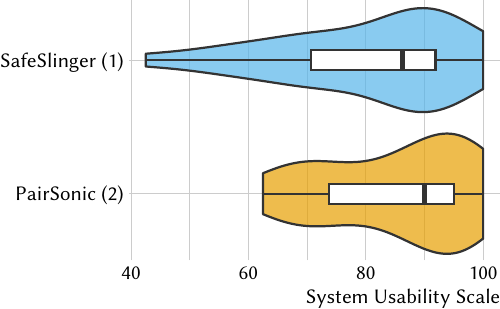}
    \caption
    {
    Participants who first encountered \SafeSlinger.
    }
    \Description{%
    The plot illustrates the distribution of System Usability Scale (SUS) scores for participants who first used SafeSlinger, followed by PairSonic. The plot provides a visual representation of the density of these scores, which is wide at the top, indicating a larger concentration of higher SUS scores. Below this, PairSonic's violin plot and boxplot indicate a slightly more narrow distribution, primarily concentrated in the upper range of the SUS score spectrum, suggesting a slightly higher overall usability rating for PairSonic when it's used after SafeSlinger. 
	}
    \label{fig:evaluation-sus-slinger-first}
\end{subfigure}%
\begin{subfigure}{0.5\columnwidth}
    %\centering
		\includegraphics[scale=0.78]{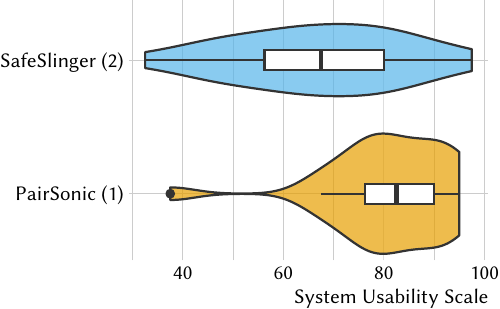}
    \caption
    {
    Participants who first encountered \OurApproach.
    }
    \Description{%
    The plot shows the SUS scores for participants who initially interacted with PairSonic before using SafeSlinger. The distribution shows that participants' SUS scores for PairSonic have a high density in the upper range, reflected in the violin plot's bulge. Conversely, SafeSlinger's scores are distributed across a broader range, with a lower median score, which can be seen in the spread of the boxplot.
	}
    \label{fig:evaluation-sus-acoustic-first}
\end{subfigure}
\caption%
{
Comparison of our participants' \gls{SUS} scores for \SafeSlinger and \OurApproach, depending on the order they encountered both systems.
The violin plots show a density estimation of the distributions. The boxplots show quartiles, median, and outliers.
}
\label{fig:evaluation-sus-depending-order}
\Description{%
    A two-part image with two violin plots each showing the SUS scores depending on the order particpants experienced  SafeSlinger or PairSonic first. Both plots show the distribution density, median, and IQR of scores, suggesting how initial exposure to each system may influence the user's subsequent usability ratings.
}
\end{figure}

\begin{table}[h!]
\centering
\caption[Correlations]
{
  Kendall's correlation between control and dependent variables.
}
\tablefontsize
\setlength{\tabcolsep}{7.5pt}
\begin{tabularx}{\columnwidth}{@{} r X rrrrrrrrrr @{}}
\toprule
&&
\multicolumn{1}{c}{\tableheadline{2}} &
\multicolumn{1}{c}{\tableheadline{3}} &
\multicolumn{1}{c}{\tableheadline{4}} &
\multicolumn{1}{c}{\tableheadline{5}} &
\multicolumn{1}{c}{\tableheadline{6}} &
\multicolumn{1}{c}{\tableheadline{7}} &
\multicolumn{1}{c}{\tableheadline{8}} &
\multicolumn{1}{c}{\tableheadline{9}} &
\multicolumn{2}{c}{\tableheadline{ATI}} \\
\midrule
1 & Smartphone Familiarity &  -.15 & -.1 &  .19 &  .11 &  .15 & .16 & -.09 & .2 & -.14 &\\
\rowcolor{\altrowcolor} 2 & Order (\OurApproach First) && .11 &  \ssig{-.33} &  -.16 &  -.15 & -.11 & .11 & -.24 & -.18 &\\
3 & Preference (\OurApproach) &&&  \sig{-.33} &  .19 &  -.12 & 0 & .14 & .12 & .18 &\\
\rowcolor{\altrowcolor} 4 & SUS \SafeSlinger &&&&  .17 &  \sig{.29} & \sig{.24} & \ssig{-.34} & 0 & .07 &\\
5 & SUS \OurApproach &&&&&  \ssig{.31} & \sig{.28} & -.12 & -.06 & .16 &\\
\rowcolor{\altrowcolor} 6 & Security \SafeSlinger &&&&&& \sig{.3} & 0 & -.05 & .09 &\\
7 & Security \OurApproach &&&&&&& -.15 & .04 & .14 &\\
\rowcolor{\altrowcolor} 8 & Time \SafeSlinger &&&&&&&& .09 & .02 &\\
9 & Time \OurApproach &&&&&&&&& .14 &\\
\bottomrule
\tablenotesx{12}{%
\textit{Note:} $N=45$\quad
\pnote \,$p<.05$\quad
\ppnote \,$p<.01$
}
\label{tab:correlations}
\end{tabularx}
\end{table}

\section{Questionnaire}\label{sec:questionnaire}
We provide a translation of our questionnaire.
We gave the questionnaire to the participants in the native language of the country where we ran the study.

\subsection{Experiment A}
\begin{task}
Please indicate the degree to which you agree/disagree with the following statements.
\end{task}

\rating{Five responses ranging from \textquote{Strongly disagree} to \textquote{Strongly agree}}

\begin{enumerate}
    \item I think that I would like to use this system frequently.
    \item I found the system unnecessarily complex.
    \item I thought the system was easy to use.
    \item I think that I would need the support of a technical person to be able to use this system.
    \item I found the various functions in this system were well integrated.
    \item I thought there was too much inconsistency in this system.
    \item I would imagine that most people would learn to use this system very quickly.
    \item I found the system very cumbersome to use.
    \item I felt very confident using the system.
    \item I needed to learn a lot of things before i could get going with this system.
    \item I think that this system is secure.
\end{enumerate}

\subsection{Experiment B}
\rating{Same content as experiment A}

\subsection{Preference}
\begin{enumerate}
    \item Which system did you like better? \rating{Experiment A / Experiment B}
\end{enumerate}

\subsection{Groups}
\begin{task}
Which of the types of digital communication groups listed below do you participate in?
What is the average number of participants in these groups?
For which of these groups do you want to know exactly who is part of the group  or that no unauthorized person  has access?
\end{task}

\rating{For each of the following categories: a checkbox \textquote{Participation?}, a field \textquote{Size?}, a checkbox \textquote{Security?}.}

\begin{enumerate}
    \item Public chat group (e.\,g., Discord, Telegram, IRC)
    \item Private chat group (e.\,g., WhatsApp, Signal)
    \item Professional chat group (e.\,g., Microsoft Teams)
    \item Online forum
    \item Group in social network (e.\,g., Facebook group)
    \item Audio/video conference (e.\,g., TeamSpeak, Zoom)
    \item Mailing list
    \item Collaboration tools (e.\,g., Google Docs, Etherpad, Miro)
    \item Project planning tools (e.\,g., Jira, Trello, GitHub)
    \item \rating{Additional free text fields}
\end{enumerate}

\subsubsection{Total Number of Groups}
\begin{enumerate}
    \item Please estimate the total number of digital communication groups you participate in. \rating{Number field}
\end{enumerate}

\subsection{Demographic Information}
\begin{enumerate}
    \item Please state your gender. \rating{Male / Female / Diverse / Free text / No answer}
    \item Please state your field of activity or field of studies. \rating{Free text response}
    \item How old are you? \rating{18--19 / 20--24 / 25--29 / 30--34 / 35--39 / 40--44 / 45--49 / 50--54 / 55--59 / 60--64 / 65--69 / 70 or older / No answer}
    \item Please state your general education. \rating{Still in school / Lower secondary education  / Polytechnic high school / Intermediary secondary education / University entrance qualification / No general school leaving certificate / Free text / No answer}
    \item Please state your professional/vocational education. \rating{ Vocational training or training in dual system / Technical college diploma / Technical college diploma in the former GDR / Bachelor / Master / Diploma / PhD / No professional or vocational degree / Free text / No answer}
    \item Are you currently a student enrolled in a degree program (Bachelor, Master, Diploma, State examination, Magister)? \rating{Yes / No / No answer}
    \item Have you been using a smartphone for more than two years?  \rating{Yes / No / No answer}
\end{enumerate}

\subsection{Technology}
\begin{task}
In the following questionnaire, we will ask you about your interaction with technical systems.
The term \textquote{technical systems} refers to apps and other software applications, as well as entire digital devices (e.\,g., mobile phone, computer, TV, car navigation). Please indicate the degree to which you agree/disagree with the following statements.
\end{task}

\rating{Six responses from \textquote{Completely disagree} to \textquote{Completely agree}}

\begin{enumerate}
    \item I like to occupy myself in greater detail with technical systems.
    \item I like testing the functions of new technical systems.
    \item I predominantly deal with technical systems because I have to.
    \item When I have a new technical system in front of me I try it out intensively.
    \item I enjoy spending time becoming acquainted with a new technical system.
    \item It is enough for me that a technical system works; I don't care how or why.
    \item I try to understand how a technical system exactly works.
    \item It is enough for me to know the basic functions of a technical system.
    \item I try to make full use of the capabilities of a technical system.
\end{enumerate}

\end{document}